\documentclass[10pt,conference]{IEEEtran}
\usepackage{cite}
\usepackage{amsmath,amssymb,amsfonts}
\usepackage{graphicx}
\usepackage{textcomp}
\usepackage{xcolor}
\usepackage{hyperref}       
\usepackage{url}            
\usepackage{booktabs}       
\usepackage{amsfonts}       
\usepackage{nicefrac}       
\usepackage{xspace}
\usepackage{color}
\usepackage{soul}
\usepackage{comment}
\usepackage{tabularx}
\usepackage{array}
\usepackage{multirow}
\usepackage{booktabs}
\usepackage{enumitem}
\usepackage{alltt}
\usepackage{siunitx}
\usepackage{caption}
\usepackage{algorithm}
\usepackage{algpseudocode}
\usepackage{tcolorbox}
\usepackage[numbers]{natbib}
\usepackage[para]{footmisc}
\usepackage{orcidlink}

\definecolor{cAdviceBg}{RGB}{247,250,255}
\definecolor{cAdviceFrame}{RGB}{72,116,176}
\definecolor{cBlockBg}{RGB}{255,255,248}
\definecolor{cBlockFrame}{RGB}{220,180,90}

\newtcolorbox{issuebox}{
  colback=cAdviceBg,
  colframe=cAdviceFrame,
  boxrule=0.8pt,
  arc=2pt,
  left=5pt,
  right=5pt,
  top=5pt,
  bottom=5pt
}

\newtcolorbox{plainblock}{
  colback=cBlockBg,
  colframe=cBlockFrame,
  boxrule=0.5pt,
  arc=1.5pt,
  left=5pt,
  right=5pt,
  top=4pt,
  bottom=4pt,
  before skip=4pt,
  after skip=4pt
}
\def\BibTeX{{\rm B\kern-.05em{\sc i\kern-.025em b}\kern-.08em T\kern-.1667em\lower.7ex\hbox{E}\kern-.125emX}}
    
\newcommand{\approach}{\textsc{LivePlan}\xspace}
\newcommand{\graph}{\textsc{Graphectory}\xspace}
\newcommand{\lang}{\textsc{Langutory}\xspace}
\newcommand{\monitor}{\emph{Monitor}\xspace}
\newcommand{\advisor}{\emph{Advisor}\xspace}
\newcommand{\processor}{\emph{Intervention Processor}\xspace}
\newcommand{\SA}{SWE-agent\xspace}
\newcommand{\monitoronly}{\emph{Predefined Advice}\xspace}
\newcommand{\periodicadvisor}{\emph{Periodic Advisor}\xspace}
\newcommand{\Vthree}{DeepSeek-V3\xspace}
\newcommand{\gemini}{Gemini-2.5-Flash\xspace}
\newcommand{\minimaxtwo}{MiniMax-M2.5\xspace}
\newcommand{\minimaxthree}{MiniMax-M3\xspace}
\newcommand{\gpt}{GPT-5.2-Codex\xspace}
\newcommand{\swebv}{SWE-bench Verified\xspace}
\newcommand{\swebp}{SWE-bench Pro\xspace}
\newcommand{\sweprm}{SWE-PRM\xspace}
\newcommand{\sage}{SAGE\xspace}
\newcommand{\wink}{Wink\xspace}

\newcommand{\reyhan}[1]{\textcolor{cyan}{\textbf{Reyhan:} #1}}
\newcommand{\shuyang}[1]{\textcolor{magenta}{\textbf{Shuyang:} #1}}

\IEEEoverridecommandlockouts
\begin{document}

\title{Online Monitoring and Corrective Steering of Programming Agents}


\author{
\IEEEauthorblockN{
Shuyang Liu\IEEEauthorrefmark{2},
Saman Dehghan\IEEEauthorrefmark{2},
Ji Young Kim\IEEEauthorrefmark{2},
Jatin Ganhotra\IEEEauthorrefmark{1},
Martin Hirzel\IEEEauthorrefmark{1},
and Reyhaneh Jabbarvand\IEEEauthorrefmark{2}
}
\IEEEauthorblockA{
\IEEEauthorrefmark{2}University of Illinois Urbana-Champaign, USA,
\{sl225, samand2, jyk14, reyhaneh\}@illinois.edu\\
\IEEEauthorrefmark{1}IBM, USA,
\{jatinganhotra, hirzel\}@us.ibm.com
}
\vspace{0.3em}
}

\maketitle

\vspace*{-40pt}
\begin{abstract}
Fixing GitHub issues in large-scale projects is a long-horizon task, especially when a fix requires changes across multiple locations or the issue description lacks the information needed to localize and repair it. As a result, agents traverse long trajectories that are prone to inefficiency and error: they drift away from their intended plan, repeat failed actions, or terminate without a working patch. This paper proposes \approach to monitor, detect, and correct such behavioral inefficiencies and drifts in real time. \approach decouples \emph{judging} from \emph{advising}: a deterministic, rule-based monitor examines general signals over the trajectory to detect issues without invoking an LLM, and only when an issue is detected does it consult an advisor LLM for a high-level, next-step correction. This design avoids the misleading re-planning and costly interventions of prior approaches. We implement \approach on top of \SA and evaluate it using five LLMs (three as executor agents and two as advisors) across \swebv and \swebp. Compared to vanilla \SA, \approach notably improves issue resolution rates, achieving consistent gains of up to 15.2\% (average: 9.9\%), while incurring only an additional cost of \$0.08 per instance. The additional solutions concentrate on \emph{medium} and \emph{hard} instances. \approach consistently outperforms alternative approaches in resolution rate, with minimal regression on already successful runs and new successes on problems that no baseline solves.

\end{abstract}



\section{Introduction}
\label{sec:intro}

LLM-based agents solve complex software-engineering tasks by following a \emph{plan}: the sequence of issue-resolution phases prescribed in their system prompt, e.g., localize, reproduce, patch, validate~\cite{SWE-agent}. Yet an agent may drift away from this plan during execution~\cite{plan-compliance}, e.g., skipping validation, or exhibit inefficiencies~\cite{liu2025empirical,trail} such as repeated actions and tool call failures, which cause failures or inflate costs even when the agent succeeds~\cite{graphectory}. Existing attempts to correct such drift suffer from three main limitations: 


    (1) They fold \emph{judging} (is the trajectory in trouble) and \emph{advising} (what should the agent do to fix the trouble) into a single mechanism: an advisor LLM serves as the judge to evaluate the trajectory and, based on that evaluation, recommends corrective actions for the remainder of the trajectory~\cite{shinn2023reflexion}. Judging and advising are inherently \emph{conflicting tasks}: an advisor prompted to diagnose and provide corrective advice is incentivized to find a problem, a tendency related to the documented sycophantic bias of LLMs toward framing their prompts~\cite{sharma2024towards}. The advisor may surface a problem that does not exist and impose misleading advice that derails a run that was, in fact, on track. \emph{Overcoming this limitation requires separating judging and advising.}
    
\begin{figure}
        \centering
        \vspace*{-30pt}
        \includegraphics[width=0.9\linewidth]{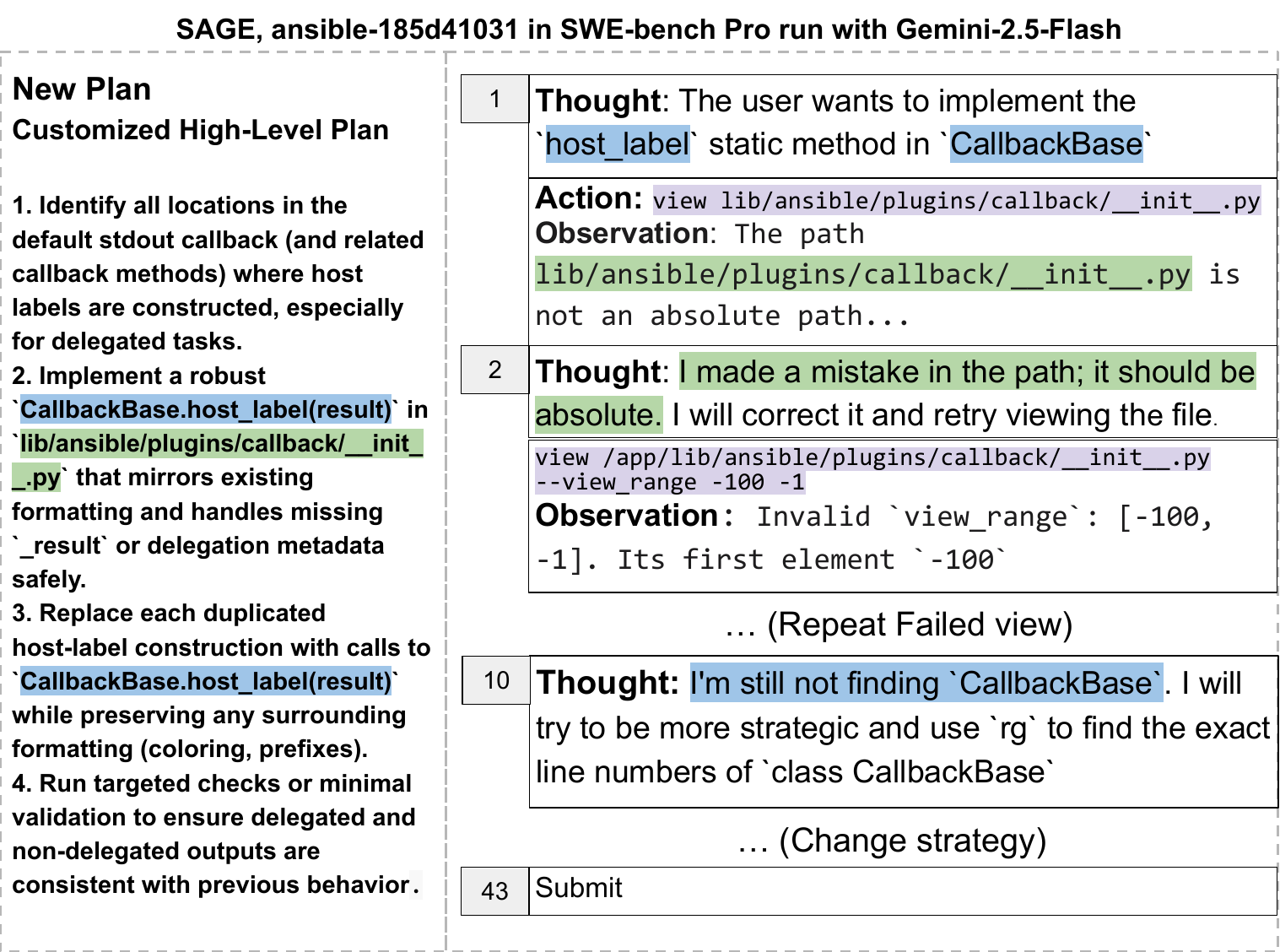}
        \vspace*{-5pt}
        \caption{An example of \sage~\cite{sage} incorrect advice (a new global plan), causing \SA to fail on a \swebp~\cite{swebenchpro} issue it had originally resolved in a vanilla run.}
    \label{fig:sage_example}
    \end{figure}

    \sloppy Figure~\ref{fig:sage_example} shows an example where \sage~\cite{sage} re-plans a trajectory for an issue originally resolved by \SA~\cite{SWE-agent} with \gemini. \sage uses \gpt as an advisor to analyze an existing trajectory---blind to the resolution status---and generates a new plan for re-execution to resolve possible trajectory problems. The advisor's plan incorrectly directs the agent to implement \texttt{\small{CallbackBase}} in \texttt{\small{lib/ansible/plugins/callback/\_\_init\_\_.py}}, which does not exist in the repository. This causes the agent to struggle through its first 10 steps, trying to locate a file that does not exist, ultimately terminating without patching.

    (2) Using an LLM-as-a-judge requires determining how often to trigger it for analysis. Some techniques wait until the end of the trajectory to perform a post-hoc analysis and rerun the agent with advice (new plan) based on the analysis~\cite{sage,trail,mast}. However, although the advisor can provide a global plan to guide the agent from the beginning, the benefit comes with caveats: if the plan is incorrect, similar to Fig.~\ref{fig:sage_example}, the agent may not be able to understand and recover, given it is instructed to follow the plan. This is also costly, as it feeds the entire trajectory to the advisor LLM and reruns the agent. Other techniques periodically ask the advisor to evaluate a sliding window of the trajectory~\cite{erdogan_et_al_2025,wink,swe_prm}. However, if the frequency is too high~\cite{erdogan_et_al_2025,wink}, the approach remains costly; if it is too low~\cite{swe_prm}, the steering feedback arrives too late, making it hard for the agent to recover from the failure mode. \emph{Overcoming this limitation entails triggering trajectory analysis only when necessary.} 
    
\begin{figure}[t]
        \centering
        \includegraphics[width=0.9\linewidth]{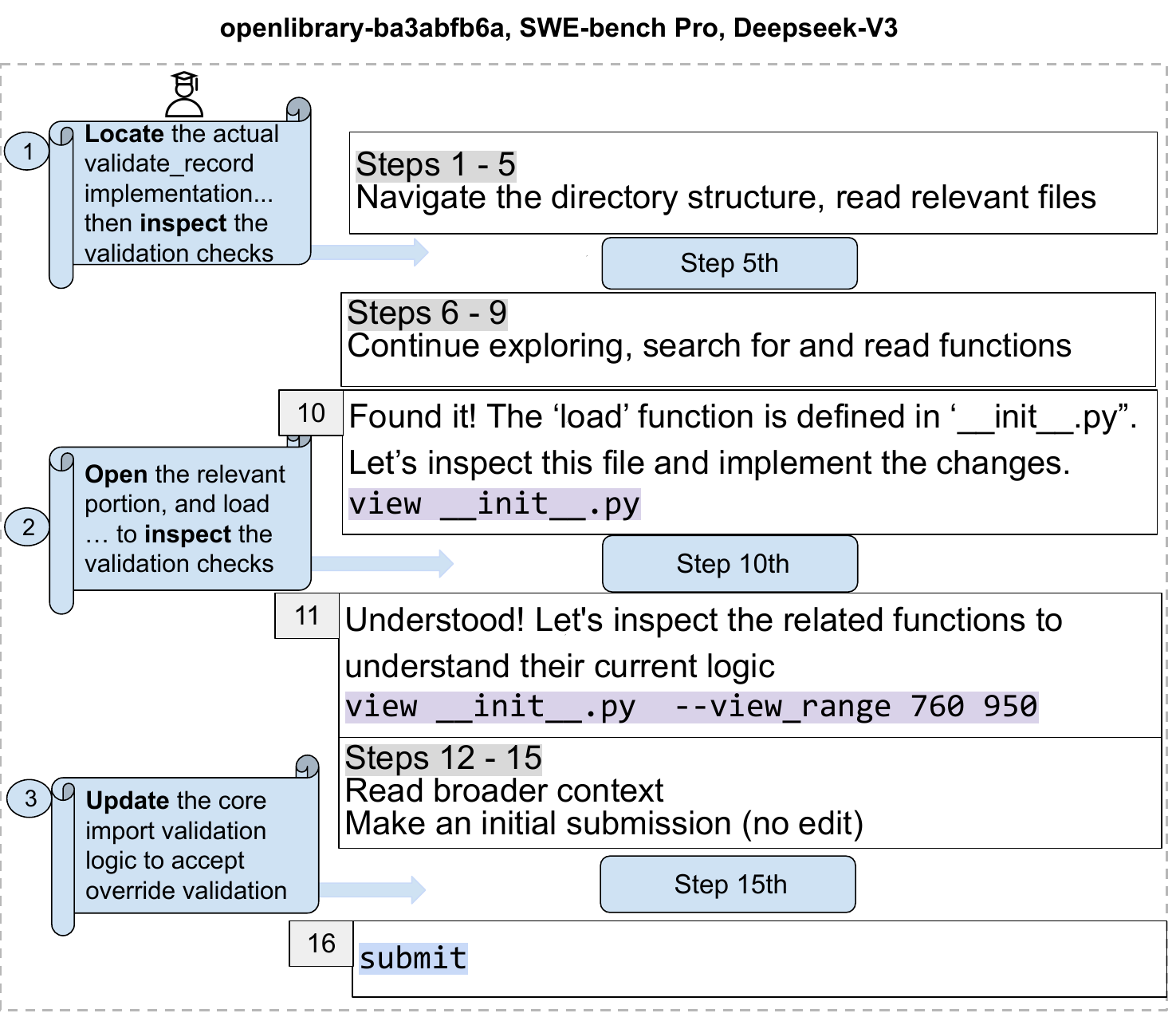}
        \caption{An advisor periodically checking the trajectory fails to resolve an issue due to late feedback.}
        \label{fig:periodic-approach-comparison-example}
    \end{figure}

    Figure~\ref{fig:periodic-approach-comparison-example} shows the execution of \SA with \Vthree on \texttt{\small{Openlibrary-ba3abfb6a}} from \swebp~\cite{swebenchpro}. A periodic advisor, similar to SWE-PRM~\cite{swe_prm}, evaluates the trajectory every five steps. During the first intervention after step~5, the executor is advised to continue exploring the repository to locate the buggy file. At step 10, it views the exact bug location. However, before taking any repair action in the next step, the advisor incorrectly concludes that the buggy location has not been found and that the agent should keep exploring, causing the agent to continue exploring without patching for the next five steps. After step 15, the advisor offers a patching suggestion, but the guidance arrives too late: the executor has already decided to terminate the run and resubmits, leaving the issue unresolved.

    (3) While LLM-as-a-judge offers generalizability, i.e., the LLM draws on its internal knowledge to flag problems that are not necessarily encoded, it comes at the cost of reliability: LLMs are known to be unreliable evaluators that both miss real problems and flag spurious ones~\cite{trail}. \emph{Overcoming this limitation entails a deterministic judge whose approach can flag a broad class of behavioral drifts.}

We propose \approach for online monitoring of agent trajectories to detect behavioral drifts and intervene with corrective steering. \approach separates \emph{judgment} from \emph{advising}: it relies on a deterministic, rule-based monitor that examines a set of \emph{general signals} over the trajectory to determine behavioral drifts (\S \ref{sec:monitor}). These signals are general rather than tied to a specific benchmark or repository, so they flag a broad class of behavioral drift. Upon detection of a behavioral drift, \approach generates predefined or custom 
high-level, corrective, next-step advice to steer the trajectory (\S \ref{sec:advisor}). By separating judgment from advising and planning, judging with a deterministic monitor that does not hallucinate a non-existent problem, and advising only when that monitor fires, \approach addresses the limitations of prior work.

We implement \approach on top of \SA~\cite{SWE-agent} and compare its performance against four baselines and ablated versions on the \swebv~\cite{swebench} and \swebp~\cite{swebenchpro} datasets, using five diverse general and reasoning LLMs (three as executors and two as advisors). The comprehensive evaluation confirms the effectiveness of \approach in consistently improving the resolution rate of \SA across all models and benchmarks, with consistent gains of up to 15.2\% (average: 9.9\%). \approach not only consistently resolves more GitHub issues than the baselines but also guides the agent more effectively toward solving \emph{medium} and \emph{hard} issues. Besides improving the overall resolution rate, \approach yields trajectories that are more compliant with the predefined instruction plan in \SA~\cite{plan-compliance}, i.e., explore, localize, repair, and validate~\cite{swe-agent-config}. 
Our contributions are:

\begin{itemize}[leftmargin=*]
    \item \textbf{Technique.} We propose \approach, an online monitoring and intervention technique that \emph{decouples} judging from advising for agent trajectory steering. \approach pairs a deterministic, rule-based monitor to detect \emph{whether and where} a trajectory is in trouble without invoking an LLM, and thus without the risk of hallucinating a non-existent problem (\S \ref{sec:monitor}). The LLM advisor is consulted only when the monitor fires and only for a high-level, next-step correction (\S \ref{sec:advisor}).
    
    \item \textbf{Empirical Evaluation.} We perform a large-scale evaluation to assess the effectiveness of \approach and other techniques on two widely used GitHub issue datasets. The study results in \num{7752} trajectories, \num{4668} of which were generated by intervention approaches, providing a valuable dataset for future research. We conducted a manual analysis of successful and unsuccessful interventions by \approach and other baselines, discussing major root causes for intervention failures and guidelines for future research. 
    
\end{itemize}

\vspace{-3pt}
\section{\approach Approach}
\label{sec:study-design}

\begin{figure}
    \centering
    \includegraphics[width=0.9\columnwidth]{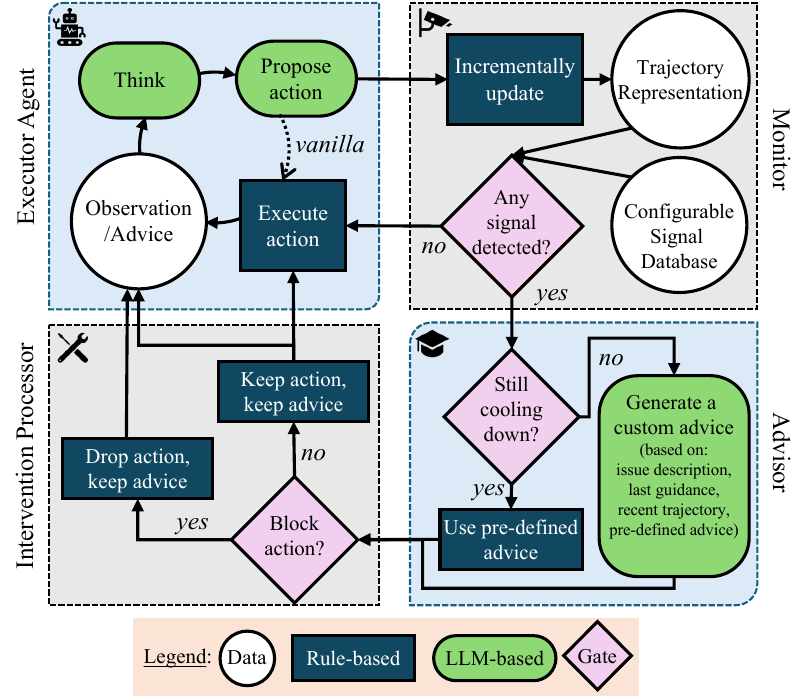}
    \caption{Overview of \approach workflow.}
    \label{fig:planner-arch}
\end{figure}

\begin{table*}[t]
\centering
\small
\renewcommand{\arraystretch}{0.92}
\setlength{\tabcolsep}{8pt}
\setlength{\aboverulesep}{2pt}
\setlength{\belowrulesep}{2pt}
\caption{Blocking and non-blocking behavioral drifts detected by \approach.}
\label{tab:monitor-rules}
\begin{tabular}{llllp{5.4cm}}
\toprule
\textbf{Type} &
\textbf{Category} &
\textbf{Signal} &
\textbf{Behavioral Drift} &
\textbf{Description} \\
\midrule
\multirow{5}{*}{Blocking}
& \multirow{3}{*}{Plan Violation}
& \multirow{3}{*}{\centering\parbox{2.5cm}{Skipping phase(s) in \lang.}}
& Premature Patching
& Patch before Localization. \\
& & & Skip Patching
& Finalize without producing a patch. \\
& & & Skip Validation
& Submit before validating the patch. \\
\cmidrule(lr){2-5}
& \multirow{2}{*}{Oscillation}
& \multirow{2}{*}{\centering\parbox{2.8cm}{Repeated back-edges in \graph}}
& Thought Oscillation
& Repeated reasoning pattern without meaningful progress. \\
& & & Action Oscillation
& Repeated execution loops, such as self-loops or cyclic action sequences. \\
\midrule
\multirow{5}{*}{Non-blocking}
& \multirow{4}{*}{Long Stagnation}
& \multirow{4}{*}{\centering\parbox{2.5cm}{Long same-phase sequence in \lang}}
& Prolonged Navigation
& Spends many steps in navigation. \\
& & & Prolonged Reproduction
& Prolonged bug reproduction. \\
& & & Prolonged Patching
& Performs many consecutive edits. \\
& & & Prolonged Validation
& Spends excessive effort validating. \\
\cmidrule(lr){2-5}
& Repeated Action
& \centering\parbox{2.5cm}{A back edge in \graph}
& Action Revisit
& Repeatedly performs the same action, such as revisiting previously viewed locations. \\
\bottomrule
\end{tabular}
\vspace{-8pt}
\end{table*}

\approach consists of \emph{three} components (Figure~\ref{fig:planner-arch}): (1)~\monitor incrementally constructs process-centric representations of the trajectory, evaluates them against a configurable set of general signals to determine behavioral drifts, and invokes the \advisor upon drift detection (\S \ref{sec:monitor}); (2)~\advisor returns a predefined advice corresponding to the drift, or prompts an LLM to generate a custom advice based on issue description, trajectory context, \monitor status, and latest advice, if available (\S \ref{sec:advisor}); depending on the latest action and \monitor observation, (3) \processor interacts with the executor agent for corrective steering (\S \ref{sec:processor}).

\subsection{\monitor}
\label{sec:monitor}

Algorithm~\ref{alg:liveplan} shows the workflow of \monitor along with other components. At each step, after the executor reasons about the next action but before executing it (Line~7), \monitor tentatively updates the trajectory representations and checks them for behavioral drift (Lines~8--9). If none is observed, execution proceeds (Lines~10--13). Otherwise, \monitor consults the \advisor for corrective steering advice (Lines~14--22).

A common strategy is to feed an LLM the entire trajectory, or a slice of it, to review and surface issues. Using an LLM as a judge is prone to hallucination and error~\cite{trail,sharma2024towards}. More critically, when a single LLM performs both judging and advising, its steering suggestion is subject to its own biased assessment~\cite{lin2025llm,deng2025can,liu2026agenthallu,sharma2024towards}. To mitigate this, recent work proposes rubric-based judging, directing the LLM to score trajectories against predefined signals~\cite{rao2026autorubric}. While this constrains the judge and reduces hallucination and bias, it does not eliminate the deeper problem: evaluation still depends on the LLM applying the rubric correctly and, crucially, \emph{deterministically}. Layering a non-deterministic judge onto an already non-deterministic agent thus compounds rather than contains this instability. Also, a single-prompt rubric is often insufficient: reliable judging may require a tool-equipped sub-agent to actively inspect the environment, substantially increasing cost~\cite{chen2026unlocking}.

\begin{algorithm}[t]
\caption{\approach Runtime Monitoring and Intervention}
\label{alg:liveplan}
\footnotesize
\begin{algorithmic}[1]
\Require Issue description $I$, advisor cooldown threshold $\theta_c$, max consecutive blocking interventions $\theta_i$, long stagnation threshold $\theta_p$
\State $T \gets \emptyset$ \Comment{committed trajectory}
\State $H \gets \Call{BuildInitialHistory}{I}$ \Comment{executor-visible history}
\State $G \gets \emptyset$; $L \gets \emptyset$ \Comment{\graph and \lang monitor states}
\State $lastTrigger \gets 0$; $lastAdvisor \gets 0$; $latestAdvice \gets \emptyset$
\State $blockCount \gets 0$ \Comment{consecutive blocking interventions}

\While{task not finished}
    \State $step \gets \Call{ProposeStep}{H}$

    \State $(G',L') \gets \Call{ExtendMonitorState}{G,L,step}$ 
    \State $rules \gets \Call{CheckRules}{G',L', \theta_p}$

    \If{$rules = \emptyset$}
        \State $(T,H) \gets \Call{ExecuteAndRecord}{step,T,H}$
        \State $(G,L) \gets (G',L')$
        \State $blockCount \gets 0$
    \Else
        \State $preDefinedAdvice \gets \Call{MonitorAdvice}{rules}$
        \State $advice \gets preDefinedAdvice$

        \If{$|T| - lastAdvisor \ge \theta_c$}
            \Comment{advisor cooldown elapsed}
            \State $recentSteps \gets \Call{StepsSince}{T,lastTrigger}$
            \State $customAdvice \gets \Call{CallAdvisor}{}$
            \State $advice \gets customAdvice$
            \State $lastAdvisor \gets |T|$
        \EndIf

        \If{\Call{ContainBlockingRule}{rules} \textbf{and} $blockCount < \theta_i$}
            \State $H \gets \Call{AppendAdvice}{H,advice}$
            \Comment{do not execute action}
            \State $blockCount \gets blockCount + 1$
            \Comment{rollback: discard $G',L'$}
        \Else
            \State $(T,H) \gets \Call{ExecuteAndRecord}{step,T,H}$
            \State $(G,L) \gets (G',L')$
            \State $H \gets \Call{AppendAdvice}{H,advice}$
            \State $blockCount \gets 0$
        \EndIf

        \State $latestAdvice \gets advice$
        \State $lastTrigger \gets |T|$
    \EndIf
\EndWhile

\Function{CallAdvisor}{}
    \State \textbf{Input:} $I$, $recentSteps$, $latestAdvice$, $preDefinedAdvice$
    \State \textbf{Return:} one LLM-generated next-step advice
\EndFunction

\end{algorithmic}
\end{algorithm}









\begin{figure*}
    \centering
    \includegraphics[width=0.88\linewidth]{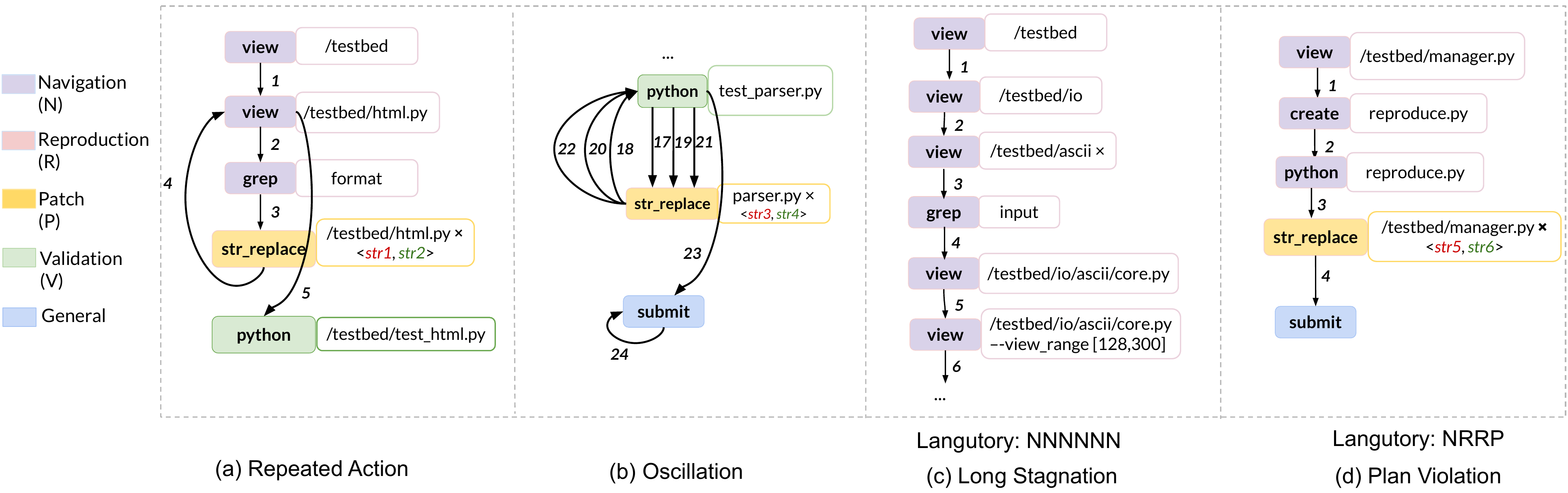}
    \caption{Examples of behavioral drifts captured by \graph and \lang.}
    \vspace{-14pt}
    \label{fig:drift-examples}
\end{figure*}

\monitor uses two process-centric trajectory representations, \graph and \lang~\cite{graphectory}, to generically encode behavioral drifts and detect them via deterministic algorithms. \graph converts the trajectory into an enriched graph, with nodes as distinct agent actions and edges as chronological execution order; \lang abstracts low-level actions into a sequence of problem-solving phases. The original \graph treats only actions as nodes, with thought included as a node property. \approach augments \graph with thought nodes and edges to also monitor ineffective thinking that manifests as behavioral drift. It also splits the \graph's \emph{localization} phase into \emph{navigation} (locating the potential bug) and \emph{reproduction} (confirming it), extending both \graph nodes and \lang alphabets with these new phases. This lets \monitor and \advisor detect behavioral drifts and issue corrective steering more specifically.

We define behavioral drift as a condition in a trajectory that deviates from common expectations. Under this definition, deviation from the specified plan, repetitive thoughts or actions that fail to advance the agent, and stagnation in any problem-solving phase are behavioral drifts. While signals defined this way are general, they may be incomplete. To account for that, \approach maintains a configurable signal database that users can augment. Furthermore, as we discuss in \S \ref{sec:advisor}, the LLM \advisor can still use its general knowledge to guide corrective steering. 

Some behavioral drifts are alarming, e.g., skipping critical problem-solving phases, such as patch validation before submission, which may result in an incorrect patch. Therefore, \emph{the execution of actions that manifest such drifts should be blocked}. Others represent inefficiencies that can be improved but are not necessarily harmful. For example, prolonged navigation through the repository to localize a bug could be due to ineffective reasoning/tooling, or a legitimate need to gather sufficient context to repair a multi-hunk bug. Therefore, execution of the corresponding actions can be allowed, accompanied by corrective steering advice. 

Table~\ref{tab:monitor-rules} lists \emph{ten} blocking and non-blocking behavioral drifts that \monitor can detect. They can be detected in \graph and \lang through generic signals:  
a back-edge in \graph, regardless of the nature of the action, indicates \emph{Repeated Action} (Figure~\ref{fig:drift-examples}a); repeated self-loops or multi-node cycles in \graph indicate \emph{Thought or Action Oscillation} (Figure~\ref{fig:drift-examples}b); a sequence of the same phase in \lang, whose length exceeds the stagnation threshold $\theta_p$, indicates \emph{Long Stagnation} (Figure~\ref{fig:drift-examples}c); and a missing phase symbol in \lang indicates skipping a plan phase, hence \emph{Plan Violation} (Figure~\ref{fig:drift-examples}d).

\vspace{-2pt}
\subsection{\advisor}
\label{sec:advisor}

\approach either provides predefined advice or invokes the LLM-based \advisor to generate custom advice (Lines~14--22). Predefined advice offers only high-level, problem-agnostic guidance; Figure~\ref{fig:predefined_messages} shows the predefined messages for \emph{Prolonged Navigation} and \emph{Action Oscillation} drifts. Although relevant to the detected drift, it may be too general to help the executor recover, so \advisor is also equipped with an LLM to generate advice specific to the problem.

\begin{figure}[t]
\centering
\small
\smallskip

\begin{plainblock}
\vspace{-5pt}
\textbf{Long Stagnation.}
\textit{Phase-specific messages are used for Navigation, Reproduction, Patching, and Validation.}
\begin{itemize}[leftmargin=1.2em, itemsep=2pt, topsep=3pt]
    \item \textbf{Prolonged Navigation.} You have the following options to explore next, which increase your chance to solve the problem:
    (1) keep exploring the code base as you do, but focus on the relevant code only;
    (2) create and run a reproduction test to better localize the bug;
    (3) if you already have enough information, edit the source code to implement the fix.
    \vspace{-5pt}
\end{itemize}
\end{plainblock}
\vspace{-7pt}
\begin{plainblock}
\vspace{-5pt}
\textbf{Oscillation.}
You are repeating action \texttt{\{action\}} in the last \texttt{\{X\}} trajectory
steps. The observation has likely not changed. Concretely reason if this action can help you resolve the issue. If not, think about a better action in the next step so that you can resolve the problem.
\vspace{-5pt}
\end{plainblock}

\caption{Examples of predefined advice in \approach.}
\vspace{-8pt}
\label{fig:predefined_messages}
\end{figure}

When invoked, the LLM-based \advisor receives four inputs: the issue description, the committed trajectory slice since the last advice (to minimize context), the latest advice if available, and the predefined advice for the detected drift. The predefined advice serves as a high-level hint to help it focus, while the recent trajectory remains the primary evidence for grounding the recommendation. \advisor outputs a next-step recommendation, rather than a long-horizon plan, keeping the intervention focused, avoiding over-constraining the executor, and preventing context clutter and degradation~\cite{liu2024lost,wang2026long,contextrot}. Since even minimal context can increase cost and latency, \approach implements a cooling mechanism: \advisor invokes the LLM only if at least $\theta_c$ trajectory steps have elapsed since the last LLM-generated advice (Line~17).

\vspace{-3pt}
\subsection{\processor}
\label{sec:processor}

The \processor component takes the advice from \advisor and, depending on the drift type, performs one of two actions: (1) for a \emph{blocking} drift, it rejects the culprit action before execution, discards the tentative \graph and \lang updates, and appends the advice to the executor history for corrective steering, up to $\theta_i$ consecutive times to avoid stalling execution (Lines~23--25); (2) for a \emph{non-blocking} drift, or once the blocking-intervention limit is reached, it lets the executor proceed and appends the advice to steer subsequent decisions (Lines~26--31).

\section{Experimental Evaluation}
\label{sec:evaluation}

\approach is pluggable to any open-source ReAct-based~\cite{yao2022react} agent. We build \approach on top of \SA as it offers a stable, well-tested implementation. 

\noindent \textbf{Benchmarks.} We evaluate on two repository-level issue resolution benchmarks: \emph{\swebv}~\cite{swebench} and \emph{\swebp}\footnote{To account for the cost, we use \swebp-Python.}~\cite{swebenchpro}, a more challenging benchmark designed to be resistant to contamination.

\textbf{Baselines.} We compare \approach against (1)~\SA using its original, predefined plan with the executor LLM, and (2)~\emph{SAGE}~\cite{sage}, which replaces this original plan with one generated by the advisor LLM given a previous trajectory. We also evaluate two ablated variants of \approach: (3) \periodicadvisor, which removes \monitor so \advisor periodically evaluates and intervenes on the trajectory, similar to \sweprm~\cite{swe_prm} and \wink~\cite{wink}\footnote{These techniques have no runnable artifacts available, so this ablation serves as the closest reproducible analog.}, and (4) \monitoronly, which removes \advisor and monitors the trajectory using only pre-defined message interventions, similar to \cite{graphectory}\footnote{Artifacts for \cite{graphectory} exist but do not implement all \approach monitoring rules and signals; we use our ablated version instead to isolate the impact of the intervention technique.}.

\textbf{LLMs.} We use three representative LLMs as executors: \emph{\Vthree}~\cite{deepseek_v3}, an open-source general-purpose model; \emph{\gemini}~\cite{gemini_2.5_flash}, a lightweight reasoning model; and \emph{\minimaxtwo}~\cite{minimax_m2.5}, a model optimized for coding and agentic workflows. Each executor is paired with a stronger but affordable advisor: \gpt~\cite{gpt_5.2_codex} (medium reasoning effort) pairs with the general-purpose executors \Vthree and \gemini, while \minimaxthree~\cite{minimax_m3} pairs with the coding-specialized \minimaxtwo. A stronger advisor is a prerequisite for meaningful intervention: an advisor no more capable than the executor offers no guidance beyond what the executor could already reach through its own reasoning. We apply the \emph{same} executor-advisor pairing across all intervention approaches, so that differences in resolution rate reflect the intervention mechanism rather than advisor capability.

\begin{table}[t]
\caption{Effectiveness on \swebp-Python.}
\setlength{\tabcolsep}{4pt}
\renewcommand{\arraystretch}{0.9}
\setlength{\aboverulesep}{2pt}
\setlength{\belowrulesep}{2pt}
\label{tab:success-rate}
\begin{tabular}{ccllc}
\toprule
\multicolumn{2}{c}{\textbf{Model}} &
\multicolumn{1}{c}{\multirow{2}{*}{\textbf{Method}}} &
\multicolumn{1}{c}{\multirow{2}{*}{\shortstack{\textbf{Success}\\\textbf{Rate (\%)}}}} &
\multicolumn{1}{c}{\multirow{2}{*}{\shortstack{\textbf{Average}\\\textbf{Cost}}}} \\
\cmidrule(lr){1-2}
\textbf{Advisor} & \textbf{Executor} & & & \\
\midrule
- & \multirow{5}{*}{\shortstack{DeepSeek\\(V3)}}
& Vanilla & 21.76 & 0.05 \\
\shortstack{GPT-5.2} & & SAGE & 18.79 \scriptsize(-2.97) & 0.23 \\
- & & \monitoronly & 25.00 \scriptsize(+3.24) & 0.11 \\
\shortstack{GPT-5.2} & & \periodicadvisor & 28.79 \scriptsize(+7.03) & 0.12 \\
\shortstack{GPT-5.2} & & \approach & \textbf{34.09} \scriptsize(+12.33) & 0.15 \\
\cmidrule(lr){1-5}
- & \multirow{5}{*}{\shortstack{Gemini\\(2.5-flash)}}
& Vanilla & 13.17 & 0.78 \\
\shortstack{GPT-5.2} & & SAGE & 18.18 \scriptsize(+5.01) & 1.16 \\
- & & \monitoronly & 18.94 \scriptsize(+5.77) & 0.98 \\
\shortstack{GPT-5.2} & & \periodicadvisor & 26.14 \scriptsize(+12.97) & 0.84 \\
\shortstack{GPT-5.2} & & \approach & \textbf{28.41} \scriptsize(+15.24) & 1.04 \\
\midrule
- & \multirow{5}{*}{\shortstack{MiniMax\\(M2.5)}}
& Vanilla & 52.5 & 0.24 \\
\shortstack{MiniMax-M3} & & SAGE & 50.38 \scriptsize(-2.12) & 0.34 \\
- & & \monitoronly & 54.17 \scriptsize(+1.67) & 0.23 \\
\shortstack{MiniMax-M3} & & \periodicadvisor & 54.54 \scriptsize(+2.04) & 0.32 \\
\shortstack{MiniMax-M3} & & \approach & \textbf{57.95} \scriptsize(+5.45) & 0.24 \\
\bottomrule
\end{tabular}
\end{table}

\textbf{Hyperparameters.} Following the default configurations of SWE-agent and SWE-bench Pro~\cite{swe-agent-config,swebenchpro}, we use a per-instance cost limit of \$2 and a model temperature of 0 for all experiments. In the ablated \periodicadvisor variant, \advisor is triggered every five steps, consistent with the \sweprm setting for fair comparison. Similarly, we set the cooling period $\theta_c$ to five steps. For long stagnation detection, we set the threshold $\theta_p$ to seven consecutive steps in the same phase, computed from Vanilla \SA runs prior to any intervention, yielding average maximum consecutive phase lengths of 5.64 for resolved instances and 7.61 for unresolved instances across all models. This threshold thus separates normal progress from likely stagnation without being tuned to any intervention method. For blocking behavioral drifts, 
\approach allows at most five consecutive interventions ($\theta_i=5$); beyond this, the agent prevents repeated blocking from stalling execution indefinitely. We run \sage using their provided prompts and pipeline without modification.

\subsection{RQ1: Effectiveness in Improving Resolution Rate}
\label{subsec:evaluation-rq1}

We run Vanilla \SA and \approach with all advisor and executor LLMs on both benchmarks. To manage experimental costs, we run the ablated versions and \sage on \swebp only across all three executor models. We prioritize \swebp as it has emerged as the dominant benchmark for long-horizon, agentic software engineering, with \swebv increasingly retired as a primary measure of progress. Tables~\ref{tab:success-rate}-\ref{tab:main_results} summarize the results.

\subsubsection{Comparison Against Vanilla Runs} 
\label{subsub:comparison-vanilla-rq1}

\textbf{\approach consistently achieves a higher resolution rate than Vanilla \SA across both benchmarks, demonstrating the effectiveness of real-time monitoring and corrective steering}. Gains are greater on \swebp, with improvements of +12.33 and +15.24 percentage points for \emph{\Vthree} and \emph{\gemini}, respectively, versus +11.20 and +10.60 points on \swebv. Even for \emph{\minimaxtwo}, which is a strong model, \approach still improves the success rates by +5.45 points on \swebp and +5.00 points on \swebv. These gains come with only modest cost increases for \emph{\Vthree} and \emph{\gemini}, while simultaneously reducing costs for \emph{\minimaxtwo}.

\begin{table}[t]
\centering
\setlength{\tabcolsep}{6pt}
\renewcommand{\arraystretch}{0.9}
\setlength{\aboverulesep}{2pt}
\setlength{\belowrulesep}{2pt}
\caption{Effectiveness on \swebv.}
\label{tab:main_results}
\begin{tabular}{ccllc}
\toprule
\multicolumn{2}{c}{\textbf{Model}} &
\multicolumn{1}{c}{\multirow{2}{*}{\textbf{Method}}} &
\multicolumn{1}{c}{\multirow{2}{*}{\shortstack{\textbf{Success}\\\textbf{Rate (\%)}}}} &
\multicolumn{1}{c}{\multirow{2}{*}{\shortstack{\textbf{Average}\\\textbf{Cost}}}} \\
\cmidrule(lr){1-2}
\multicolumn{1}{c}{\textbf{Advisor}} & \multicolumn{1}{c}{\textbf{Executor}} & & & \\
\midrule
- & \multirow{2}{*}{\shortstack{DeepSeek\\(V3)}} & Vanilla & 38.20 & 0.04 \\
\shortstack{GPT-5.2} & & \approach & \shortstack[l]{\textbf{49.40} \scriptsize(+11.20)} & 0.08 \\
\cmidrule(lr){1-5}
- & \multirow{2}{*}{\shortstack{Gemini\\(2.5-flash)}} & Vanilla & 37.80 & 0.46 \\
\shortstack{GPT-5.2} & & \approach & \shortstack[l]{\textbf{48.40} \scriptsize(+10.60)} & 0.58 \\
\midrule
- & \multirow{2}{*}{\shortstack{MiniMax\\(M2.5)}} & Vanilla & 74.20 & 0.30 \\
\shortstack{MiniMax-M3} & & \approach & \shortstack[l]{\textbf{79.20} \scriptsize(+5.00)} & \textbf{0.25} \\
\bottomrule
\end{tabular}
\end{table}

\begin{figure}
    \centering
    \includegraphics[width=0.9\linewidth]{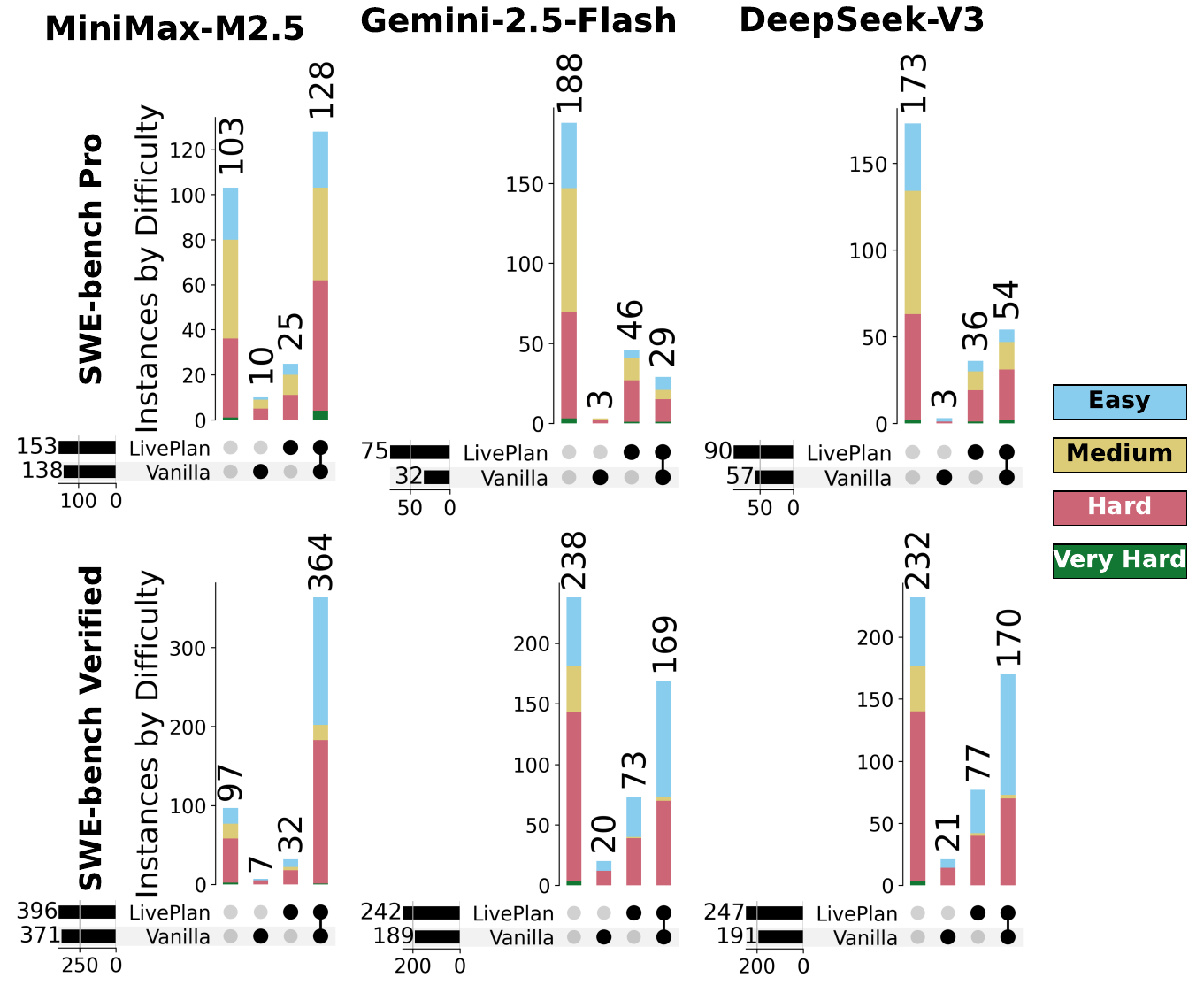}
    \caption{Resolution overlap between \approach and Vanilla.}
    \label{fig:upset_ru_difficulty}
\end{figure}

\noindent Figure~\ref{fig:upset_ru_difficulty} compares the overlap of resolved instances between \approach and Vanilla, grouped by difficulty. Black dots indicate the method(s) that resolve the corresponding instances to a bar, and gray dots mark those that do not. 
Colors indicate instance difficulty. For \swebv, we use the difficulty labels provided by the benchmark based on the estimated developer effort to resolve the issue.
For \swebp, following prior work~\cite{chen2026unlocking}, we estimate difficulty using the number of files modified in the reference patch: easy (1 file), medium (2--3 files), hard (4--10 files), and very hard ($>10$ files).
Across all model-benchmark pairs, \approach consistently resolves a substantial number of instances beyond those already solved by the vanilla. More importantly, these gains are concentrated on harder tasks: on \swebv, more than half of the instances uniquely resolved by \approach belong to the hard category; on \swebp, nearly all additional resolutions fall into the medium and hard categories. These results suggest that \textbf{online monitoring and steering are particularly beneficial for more challenging and less contamination-prone software engineering tasks.}

\subsubsection{Comparison Against \sage} 
\label{subsub:comparison-sage}

\textbf{\approach consistently outperforms \sage at even lower cost.} \sage's notable cost overhead comes from running Vanilla \SA once, feeding its entire trajectory to \advisor for re-planning, and re-executing the agent with the new plan.
\sage even underperforms the Vanilla run under \Vthree and \minimaxtwo. Although \sage produces highly specific guidance, such as exact line numbers or concrete code snippets to insert, this detail is often factually incorrect (Figure~\ref{fig:sage_example}). Executor LLMs, being generally less capable, often cannot critically evaluate or recover from such erroneous instructions and instead follow the hallucinated guidance faithfully, causing task failures.

\subsubsection{Comparison Against Ablated Versions and Similar Approaches} 
\label{subsubsec:comparison-ablation}

\begin{table}[t]
\setlength{\tabcolsep}{5.5pt}
\caption{Comparison of trigger frequency of the \advisor by different techniques.}
\label{tab:advisor_llm_trigger_stats}
\begin{tabular}{lllcc}
\toprule
\multicolumn{2}{c}{\textbf{Model}}
& \multicolumn{1}{c}{\multirow{2}{*}{\textbf{Method}}}
& \multirow{2}{*}{\shortstack{\textbf{Average}\\\textbf{Intervention}}}
& \multirow{2}{*}{\shortstack{\textbf{Trigger}\\\textbf{Rate (\%)}}} \\
\cmidrule(lr){1-2}
\textbf{Advisor} & \textbf{Executor} & & & \\
\midrule
\multirow{4}{*}{\shortstack{GPT-5.2}} & \multirow{2}{*}{\shortstack{DeepSeek\\(V3)}}
& \periodicadvisor & 3.25 & 99.6 \\
& & \approach         & 2.33 & 93.9 \\
\cmidrule(lr){2-5}
& \multirow{2}{*}{\shortstack{Gemini\\(2.5-flash)}}
& \periodicadvisor & 7.37 & 97.7 \\
& & \approach         & 2.99 & 93.9 \\
\midrule
\multirow{2}{*}{\shortstack{MiniMax\\(M3)}} & \multirow{2}{*}{\shortstack{MiniMax\\(M2.5)}}
& \periodicadvisor & 7.38 & 99.6 \\
& & \approach         & 1.47 & 83.3 \\
\bottomrule
\end{tabular}
\end{table}

\textbf{Removing \monitor or \advisor degrades the performance of \approach, demonstrating the effectiveness of both and the superiority of \approach over similar alternatives}: \sweprm (every five steps) and \wink (every step) periodically analyze the trajectory and intervene, similar to the \periodicadvisor variant. Such techniques are guaranteed to trigger the \advisor, whereas \approach does so when needed. Table~\ref{tab:advisor_llm_trigger_stats} shows that the \periodicadvisor triggers \advisor in almost all trajectories\footnote{The trajectories in which the \advisor is not triggered are shorter than the five-step threshold due to unsuccessful early termination.}, while \approach triggers it as needed. The average number of interventions in \periodicadvisor is also notably higher than \approach. As we confirm through manual analysis, the lower performance despite the higher and denser triggering rate is due to the \advisor providing misleading advice. Using \Vthree and \gemini, the average cost of \approach per instance is higher than that of \periodicadvisor, despite triggering \advisor less often. The additional cost reflects longer successful trajectories rather than more expensive interventions (details in \S \ref{subsec:evaluation-rq2}).

Compared to the \monitoronly variant, which is similar to the proposed technique in \cite{graphectory}, \textbf{\approach consistently outperforms it with a considerable margin}. This is because this variant only provides predefined, high-level messages, which may be insufficient to help the agent recover from behavioral drifts. Since \monitoronly variant detects a broader set of issues than~\cite{graphectory}, it is an upper bound on that technique's performance, i.e., \approach's margin over that technique would be even larger.

\begin{figure}
    \centering
    \includegraphics[width=0.95\linewidth]{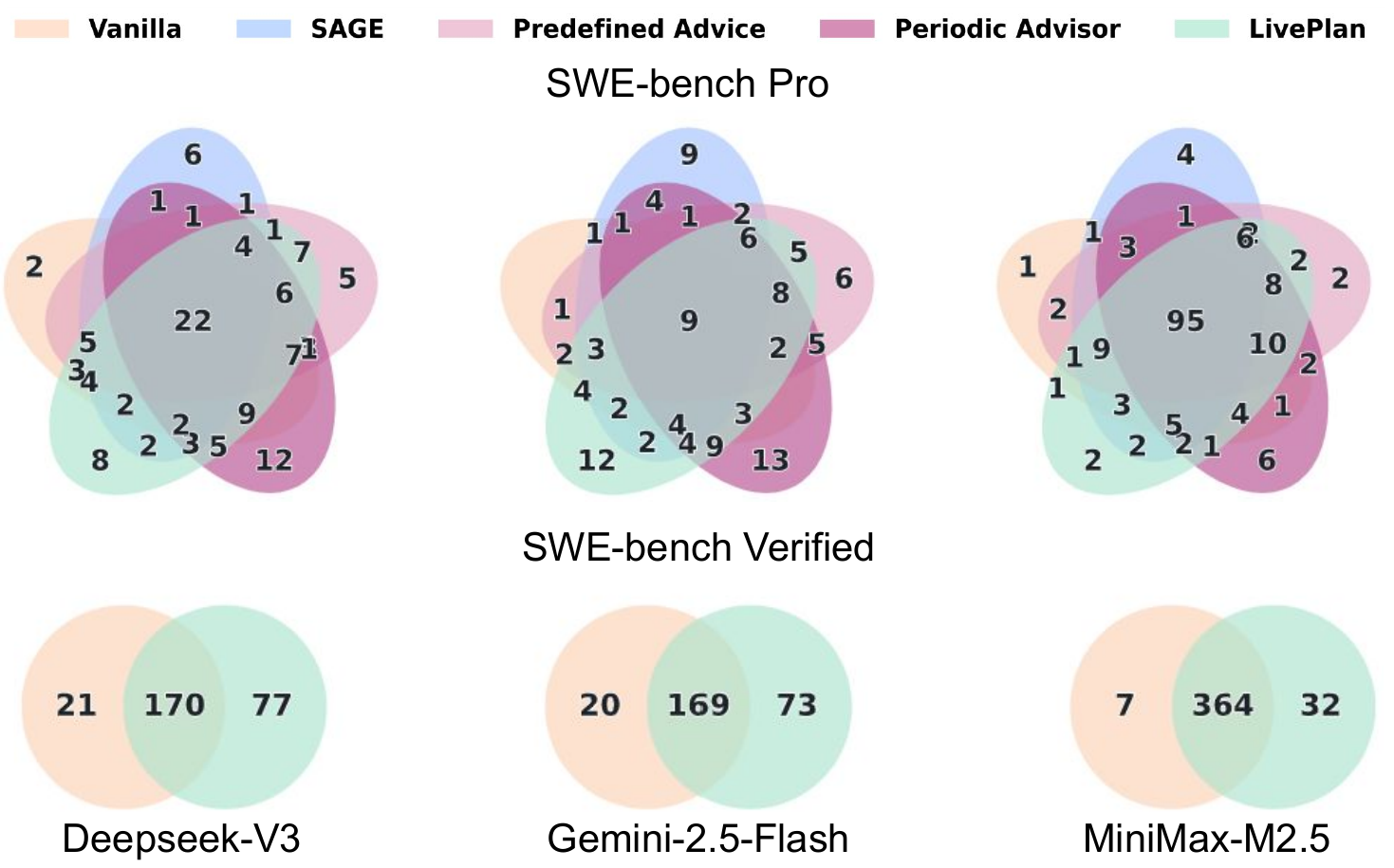}
    \caption{Overlapping resolved instances among techniques.
    }
    \label{fig:venn_resolved}
\end{figure}

Figure~\ref{fig:venn_resolved} illustrates the overlap of resolved instances by all approaches. Across most model-benchmark pairs, \approach contributes the largest set of uniquely resolved instances, e.g., 73 for \gemini and 32 for \minimaxtwo on \swebv. The distribution of resolved instances suggests that different intervention strategies exhibit complementary strengths across task subsets. \textbf{Overall, \approach offers the strongest balance between effectiveness and cost.}

\subsection{RQ2: Detailed Cost Analysis of \approach}
\label{subsec:evaluation-rq2}

\begin{table}[t]
\centering
\setlength{\tabcolsep}{1pt}
\renewcommand{\arraystretch}{0.88}
\setlength{\aboverulesep}{2pt}
\setlength{\belowrulesep}{2pt}
\caption{Cost and latency analysis. Exec: Executor; Adv: Advisor; Mon: Monitor. ``---'' indicates not applicable.}
\label{tab:cost_analysis}
\begin{tabular}{clllccccc}
\toprule
\multicolumn{3}{c}{\textbf{Setting}}
& \multicolumn{1}{c}{\multirow{2}{*}{\textbf{Method}}}
& \multirow{2}{*}{\textbf{Steps}}
& \multicolumn{2}{c}{\textbf{Cost (\$)}}
& \multicolumn{2}{c}{\textbf{Latency (s)}} \\
\cmidrule(lr){1-3}\cmidrule(lr){6-7}\cmidrule(lr){8-9}
\textbf{Dataset} & \textbf{Advisor} & \textbf{Executor}
& & & \textbf{Exec.} & \textbf{Adv.} & \textbf{Mon.} & \textbf{Adv.} \\
\midrule
\multirow{6}{*}{\rotatebox[origin=c]{90}{Pro}}
& - & \multirow{2}{*}{\shortstack{DeepSeek\\(V3)}} & Vanilla & 16 & 0.04 & --- & --- & --- \\
& GPT-5.2 & & \approach & 30 & 0.12 & 0.04 & 0.0 & 10.9 \\
\cmidrule(lr){2-9}
& - & \multirow{2}{*}{\shortstack{Gemini\\(2.5-flash)}} & Vanilla & 75 & 0.78 & --- & --- & --- \\
& GPT-5.2 & & \approach & 47 & 0.98 & 0.06 & 0.0 & 9.6 \\
\cmidrule(lr){2-9}
& - & \multirow{2}{*}{\shortstack{MiniMax\\(M2.5)}} & Vanilla & 47 & 0.24 & --- & --- & --- \\
& MiniMax-M3 & & \approach & 47 & 0.24 & 0.01 & 0.0 & 29.1 \\
\midrule
\multirow{6}{*}{\rotatebox[origin=c]{90}{Verified}}
& - & \multirow{2}{*}{\shortstack{DeepSeek\\(V3)}} & Vanilla & 21 & 0.04 & --- & --- & --- \\
& GPT-5.2 & & \approach & 24 & 0.06 & 0.02 & 0.0 & 5.8 \\
\cmidrule(lr){2-9}
& - & \multirow{2}{*}{\shortstack{Gemini\\(2.5-flash)}} & Vanilla & 52 & 0.46 & --- & --- & --- \\
& GPT-5.2 & & \approach & 38 & 0.55 & 0.04 & 0.0 & 8.1 \\
\cmidrule(lr){2-9}
& - & \multirow{2}{*}{\shortstack{MiniMax\\(M2.5)}} & Vanilla & 70 & 0.30 & --- & --- & --- \\
& MiniMax-M3 & & \approach & 58 & 0.23 & 0.02 & 0.0 & 39.5 \\
\bottomrule
\end{tabular}
\end{table}

\approach's gains come at minimal cost and overhead. Table~\ref{tab:cost_analysis} reports that across all settings, \textbf{the advisor costs only \$0.01--\$0.06 per instance on average}. \textbf{Rule-based monitoring incurs nearly zero runtime overhead}, i.e., a few milliseconds. The \advisor latency is modest relative to the runtime of long-horizon repository-level repair tasks. Beyond cost, \approach often reduces execution effort by decreasing inefficient exploration. For \gemini and \minimaxtwo, interventions generally shorten trajectories and, in some cases, even reduce the overall cost. In contrast, trajectories become longer for \Vthree because \approach discourages common shortcuts such as skipping validation~\cite{plan-compliance}, encouraging a more complete and reliable repair process. 
Regardless, the resulting cost overhead remains small.


\subsection{RQ3: Process-Centric Analysis of Trajectories}
\label{subsec:evaluation-rq3}

\begin{table*}[t]
\centering
\renewcommand{\arraystretch}{0.9}
\setlength{\aboverulesep}{2pt}
\setlength{\belowrulesep}{2pt}
\caption{Observed behavioral drifts by \approach.
}
\label{tab:trigger_distribution}
\setlength{\tabcolsep}{5pt}
\begin{tabular}{lllccccccc}
\toprule
\multicolumn{3}{c}{\textbf{Setting}}
& \multirow{2}{*}{\shortstack{\textbf{Plan}\\\textbf{Violation}}}
& \multicolumn{4}{c}{\textbf{Long Stagnation}}
& \multirow{2}{*}{\textbf{Oscillation}}
& \multirow{2}{*}{\shortstack{\textbf{Repeated}\\\textbf{Action}}} \\
\cmidrule(lr){1-3}\cmidrule(lr){5-8}
\textbf{Dataset} & \textbf{Advisor} & \textbf{Executor}
& & \textbf{Navigation} & \textbf{Reproduction} & \textbf{Patching} & \textbf{Validation}
& & \\
\midrule
\multirow{3}{*}{\shortstack{SWE-bench\\Pro}}
& \multirow{2}{*}{GPT-5.2} & \Vthree
& 80.7 & 26.1 & 1.5 & 11.4 & 11.0 & 95.5 & 28.8 \\
& & \gemini
& 43.9 & 31.8 & 5.3 & 20.5 & 20.8 & 73.5 & 59.5 \\
& MiniMax-M3 & \minimaxtwo
& 1.9 & 48.7 & 5.3 & 1.1 & 21.3 & 10.3 & 44.1 \\
\midrule
\multirow{3}{*}{\shortstack{SWE-bench\\Verified}}
& \multirow{2}{*}{GPT-5.2} & \Vthree
& 28.2 & 22.8 & 7.4 & 0.6 & 14.6 & 37.0 & 12.8 \\
& & \gemini
& 7.7 & 34.9 & 13.8 & 8.4 & 15.7 & 78.3 & 37.8 \\
& MiniMax-M3 & \minimaxtwo
& 6.2 & 42.4 & 21.2 & 0.6 & 47.8 & 6.8 & 47.0 \\
\bottomrule
\end{tabular}
\vspace{-10pt}
\end{table*}

In addition to outcome-centric analysis, we conduct three types of automated process-centric analyses to better understand the notable improvement of \approach over Vanilla.

\subsubsection{Prevalence of Triggering Signals}

\begin{figure}
    \centering
    \vspace{-10pt}
    \includegraphics[width=\linewidth]{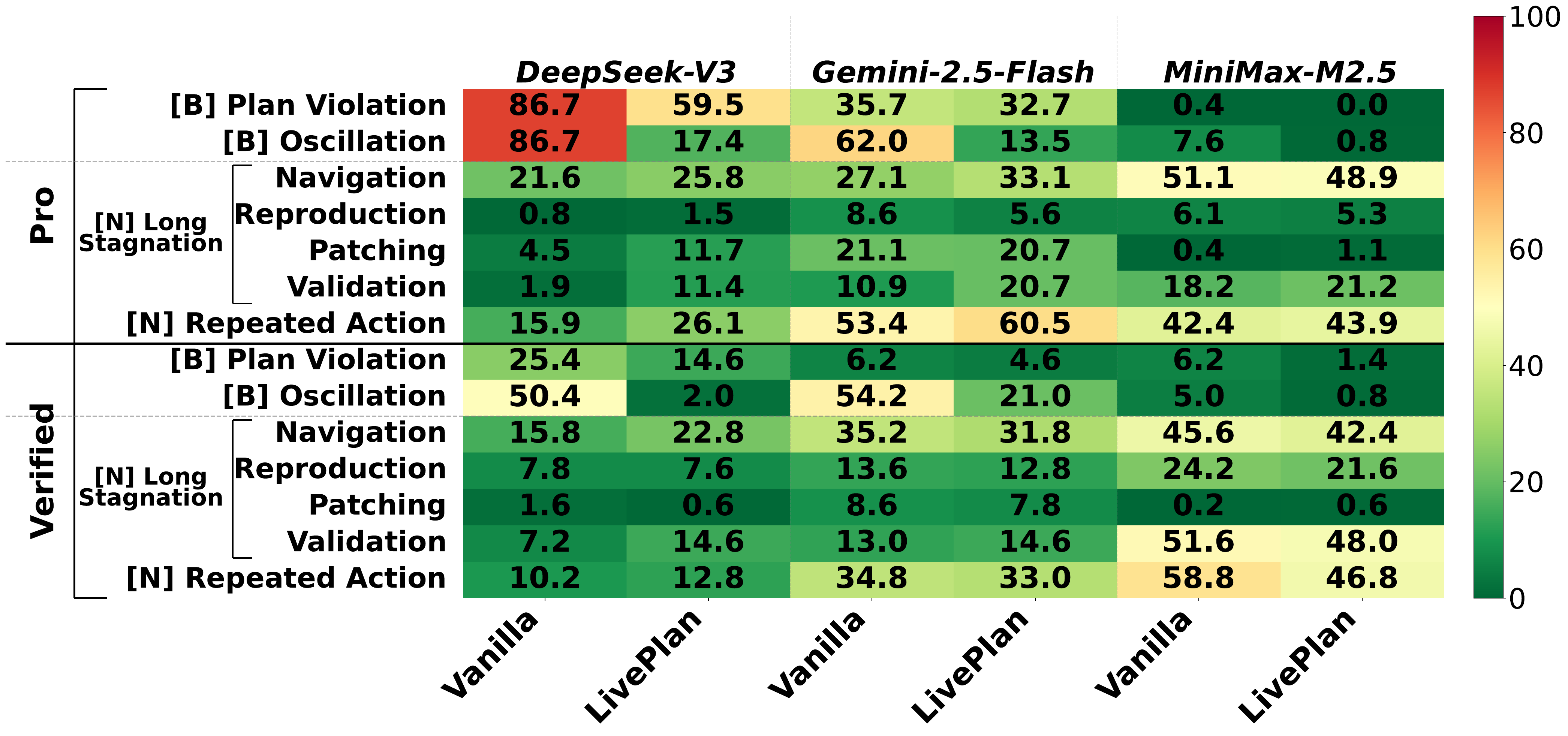}
    \caption{Behavioral drifts in final trajectories.
    }
    \label{fig:inefficiency_patterns_heatmap}
\end{figure}

We measure the prevalence of behavioral drifts in \emph{completed} trajectories of Vanilla and \approach (Figure~\ref{fig:inefficiency_patterns_heatmap}), and the percentage of instances in which \approach observes each drift during execution (Table~\ref{tab:trigger_distribution}). Drifts observed by \approach may not remain in the final trajectory, since \approach steers the agent away from them. On average, \textbf{86.16\% and 78.33\% of Vanilla trajectories in \swebp and \swebv show behavioral drifts}. They are more prevalent in \swebp, whose more complex problems require longer, drift-prone trajectories.

The exact behavioral drifts in Vanilla runs may not be observed in \approach. However, Table~\ref{tab:trigger_distribution} shows that \approach observes many similar behavioral drifts during execution and attempts to steer the agent away. Specifically, \approach significantly reduces blocking signals. \textbf{Many of the high-risk behavioral drifts, namely Plan Violation and Action/Thought Oscillation, that are prevalent in the final Vanilla trajectories and are notably observed during \approach runs do not appear in final \approach trajectories}. 55.63\% and 40.27\% of blocking behavioral drifts in Vanilla trajectories under \swebp and \swebv drops to 34.37\% and 13.67\% in \approach, respectively. 

As our manual analysis confirms (\S \ref{subsec:evaluation-rq4}), the remaining blocking drifts are due to the executor either being unable to follow the advice or choosing not to, repeating the drift after exceeding the consecutive-intervention threshold $\theta_i$. Non-blocking signals may remain in completed \approach trajectories. This is because they can signal either an inefficiency or a legitimate, prolonged attempt necessary to solve hard problems. These results show that \textbf{\approach improves execution quality by minimizing high-risk behavioral drifts while preserving the flexibility needed for long-horizon tasks.}

\subsubsection{Plan Compliance Analysis}

\begin{figure}
    \centering
    \vspace{-10pt}
    \includegraphics[width=\linewidth]{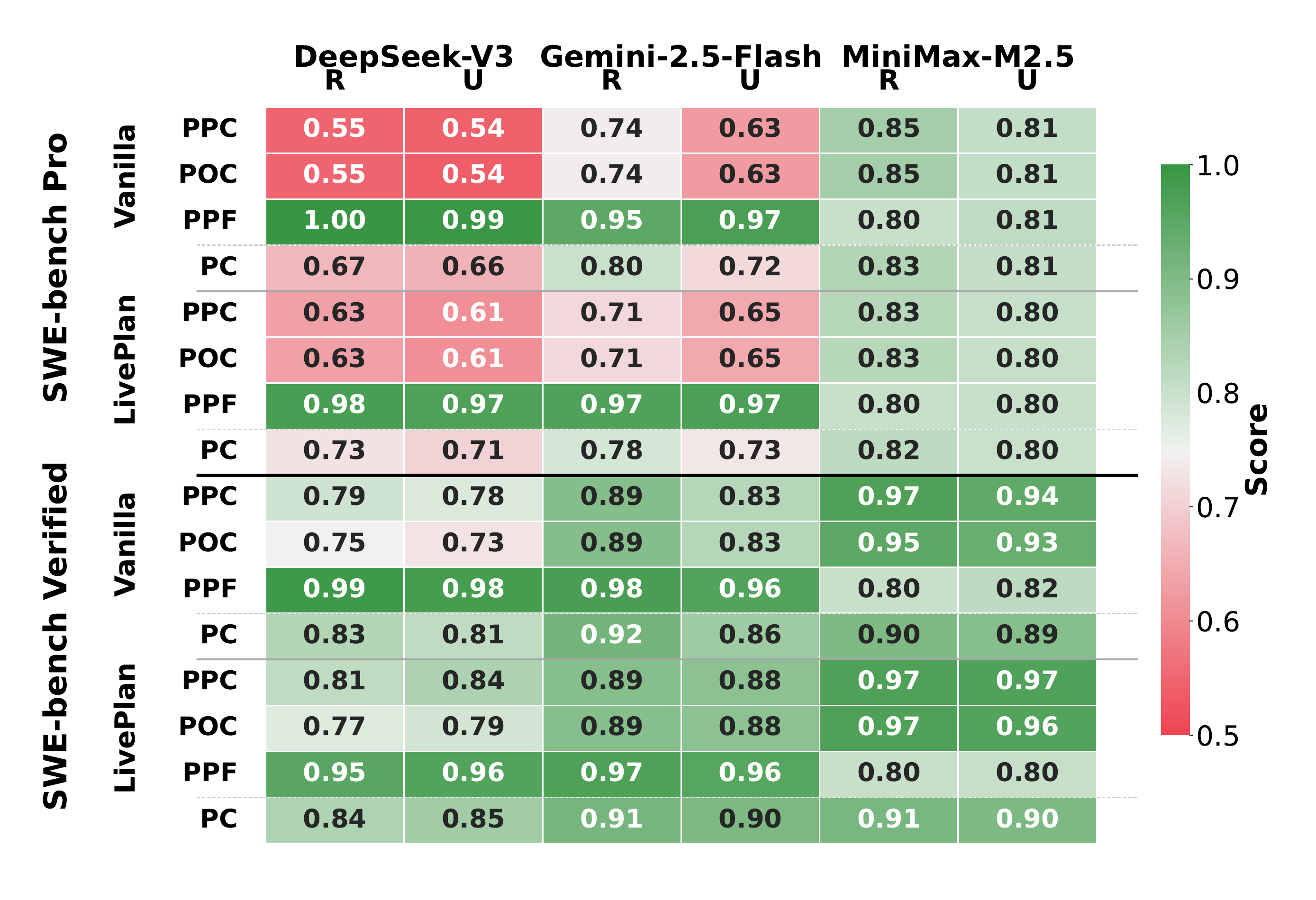}
    \vspace{-15pt}
    \caption{Average plan compliance score for Resolved (R) and Unresolved (U) instances of Vanilla and \approach.}
    \label{fig:plan_compliance_heatmap}
\end{figure}

\approach's final trajectories, specifically resolved \swebp instances by \Vthree and \gemini, exhibited Plan Violation, albeit to a lesser extent than Vanilla, motivating deeper plan-compliance analysis. Following prior work~\cite{plan-compliance}, we measure trajectory adherence to \SA's repair plan, i.e., Navigate, Reproduce, Patch, and Validate, via four metrics: \emph{Plan Phase Compliance} (PPC) penalizes skipped phases; \emph{Plan Order Compliance} (POC) penalizes out-of-order execution; and \emph{Plan Phase Fidelity} (PPF) penalizes behaviors outside the plan. The overall \emph{Plan Compliance} (PC) is the geometric mean of PPC, POC, and PPF. PC=1 only when a trajectory executes all and only the specified phases in the correct order.

Figure~\ref{fig:plan_compliance_heatmap} reports average compliance scores for resolved and unresolved instances of Vanilla and \approach.
Consistent with prior work~\cite{plan-compliance}, resolved trajectories achieve higher PC than unresolved ones, indicating that successful repairs mostly followed the intended workflow. \textbf{\approach improves PC mainly through higher PPC and POC, suggesting online intervention helps agents complete the required phases in the intended order}. PPF is occasionally lower under \approach, which we attribute to \advisor suggesting beneficial out-of-plan actions, e.g., additional regression testing to validate the patch. Since these trajectories succeed, the out-of-plan behavior appears productive rather than aimless.


\subsubsection{Strategy Divergence Analysis}

The plan-compliance analysis showed that \advisor interventions affect how closely the executor follows its prescribed plan. Independent of that plan, we next perform an apples-to-apples comparison between Vanilla and \approach trajectories to understand whether they are similar \emph{before} the intervention and how they change \emph{after} it. We compute normalized longest common subsequence (LCS) similarity between \lang of Vanilla ($X)$ and \approach ($Y$) for each instance as $\frac{2 \times |\mathrm{LCS}(X,Y)|}{|X|+|Y|}$. 
Larger values indicate more similar execution behavior. We evaluate similarity on (1) the \lang prefixes before the first intervention and (2) the \lang of complete trajectories. The former captures divergence from execution non-determinism, since both runs evolve independently before any intervention. The latter captures the combined effect of non-determinism and online intervention. 

\begin{figure}
    \centering
    \vspace{-5pt}
    \includegraphics[width=0.9\linewidth]{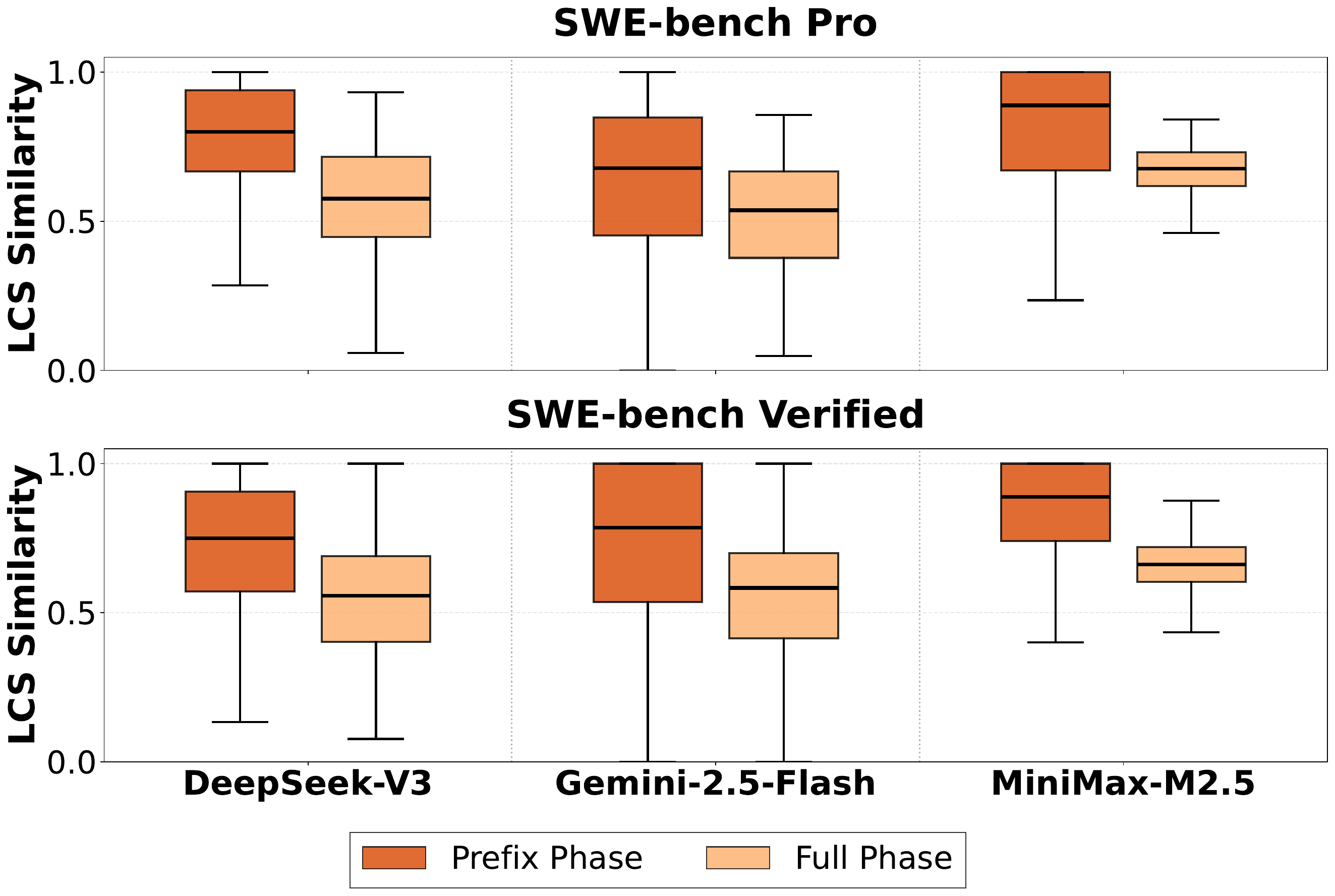}
    \caption{Trajectory similarity between Vanilla and \approach before the first intervention and over complete trajectories.}
    \label{fig:divergence_boxplot}
\end{figure}

Figure~\ref{fig:divergence_boxplot} reports the results. Prefix similarity is high, often 0.8--0.9, indicating agents initially follow similar strategies despite execution non-determinism. \textbf{After intervention, similarity consistently decreases, showing that \approach redirects the subsequent problem-solving strategy}. A paired one-sided Wilcoxon signed-rank test~\cite{wilcoxon_test} comparing prefix and full-trajectory similarity per instance confirms the significance: in every model-benchmark pair, prefix similarity is significantly higher (median paired difference $>0$, $p<1e^{-10}$). Since prefix and full-trajectory similarities are paired by instance, and difficulty is held fixed across the two measurements, the post-intervention drop reflects the intervention itself rather than task variation. While we discuss the impact of non-determinism in detail later (\S \ref{subsec:evaluation-rq5}), these results indicate that \approach's performance gains stem from targeted behavioral corrections rather than run-to-run variation.

\vspace{-8pt}
\subsection{RQ4: Analysis of Improvements and Regressions}
\label{subsec:evaluation-rq4}

\begin{table}[t]
\caption{Resolution transition matrix relative to Vanilla.
\textbf{R}: Resolved, \textbf{U}: Unresolved, \textbf{DS}: \Vthree, \textbf{GF}: Gemini-2.5-Flash, and \textbf{MM}: MiniMax-M2.5.}
\label{tab:ru_transition}
\setlength{\tabcolsep}{2pt}
\setlength{\aboverulesep}{2pt}
\setlength{\belowrulesep}{2pt}
\centerline{\renewcommand{\arraystretch}{0.9}\small\begin{tabular}{@{}ll@{ }
ccc
ccc@{}}
\toprule
\multirow{2}{*}{\textbf{Dataset}} &
\multirow{2}{*}{\textbf{Method}} &
\multicolumn{3}{c}{\textbf{R$\rightarrow$U} $\downarrow$} &
\multicolumn{3}{c}{\textbf{U$\rightarrow$R} $\uparrow$} \\
\cmidrule(lr){3-5}\cmidrule(lr){6-8}
&&
\textbf{DS} & \textbf{GF} & \textbf{MM} &
\textbf{DS} & \textbf{GF} & \textbf{MM} \\
\midrule
\multirow{4}{*}{\shortstack[l]{\swebp}}
& \sage             & 26 & 12 & 22 & 19 & 28 & 16 \\
& \monitoronly     & 16 & 13 & 15 & 25 & 28 & 15 \\
& \periodicadvisor & 16 & 11 & 17 & 35 & 42 & 22 \\
& \approach         & \textbf{2} & \textbf{2} & \textbf{7} & 33 & 38 & 21 \\
\midrule
\multirow{1}{*}{\shortstack[l]{\swebv}}
& \approach         & 12 & 7 & 7 & 56 & 68 & 28 \\
\bottomrule
\end{tabular}}
\end{table}

Table~\ref{tab:ru_transition} reports resolution transitions relative to Vanilla, computed over instances where each method is triggered. \approach consistently achieves the fewest R$\rightarrow$U regressions while producing many U$\rightarrow$R improvements\footnote{Although \periodicadvisor achieves slightly more U$\rightarrow$R transitions in some settings, it also introduces substantially more R$\rightarrow$U regressions, leading to a lower overall success rate than \approach.}, particularly on \swebp.
These results show that \textbf{\approach not only resolves notable previously unsuccessful instances but also preserves the Vanilla successes}.
We manually analyzed improvement (U$\rightarrow$R) and regression R$\rightarrow$U cases to understand root causes. Two authors \emph{independently} inspected the trajectory, intervention(s), resulting patch, and golden patch to inductively identify root causes, with categories emerging from and refined during analysis rather than defined in advance. The authors then discussed findings to reach consensus on final root-cause labels\footnote{Identified root causes and corresponding instances are available in artifacts~\cite{website}.}.

\begin{figure*}[t]
    \centering
    \includegraphics[width=0.85\linewidth]{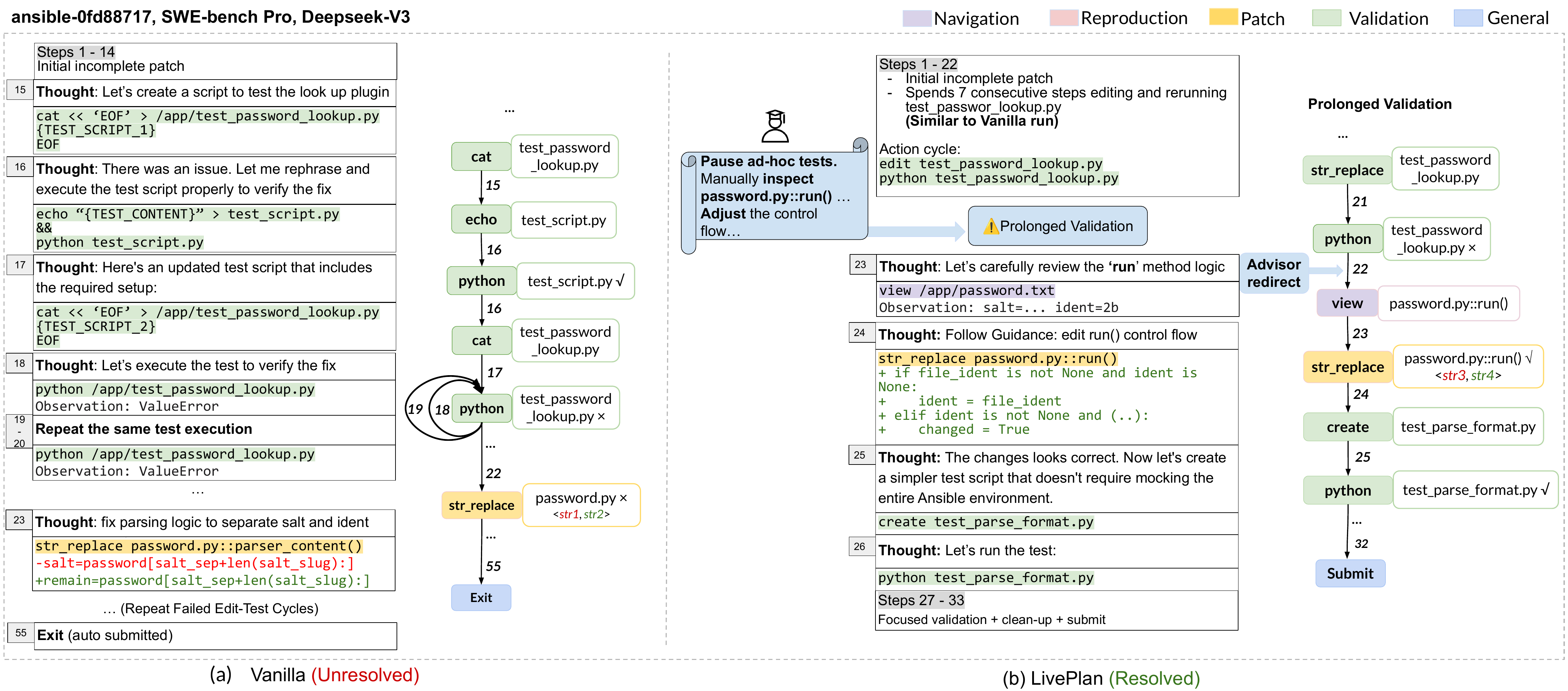}
    \caption{Case study of a U$\rightarrow$R transition. Intervention redirects the executor from ineffective validation to successful repair.}
    \vspace*{-10pt}
    \label{fig:case_study_ur}
\end{figure*}

\subsubsection{Improvements (U$\rightarrow$R)}

Our analysis identifies the following reasons for \approach's success in converting previously unsuccessful Vanilla runs: \textbf{(1) Correcting behavioral drift.} Weaker executors, particularly \Vthree and \gemini, frequently exhibit blocking behavioral drifts (Table~\ref{tab:trigger_distribution}); \approach detects them promptly and redirects execution before they persist, eliminating Oscillations in 104/128 (81.3\%) of affected trajectories and Plan Violation cases in 37/72 (51.4\%). \textbf{(2) Refocusing on the repair task.} As trajectories lengthen, executors may drift from the main repair task due to accumulated context or intermediate failures (context poisoning~\cite{chen2026unlocking}); by conditioning on the issue description and recent trajectory steps, \advisor redirects the executor away from unproductive exploration and back toward the core problem. \textbf{(3) Progressive online correction.} 182/244 (74.6\%) of U$\rightarrow$R cases receive multiple interventions, averaging 3.70 across all 244 cases, letting \approach continually correct newly emerging drifts rather than relying on a one-time redirection.

 \begin{figure*}[t]
    \centering
    \includegraphics[width=\linewidth]{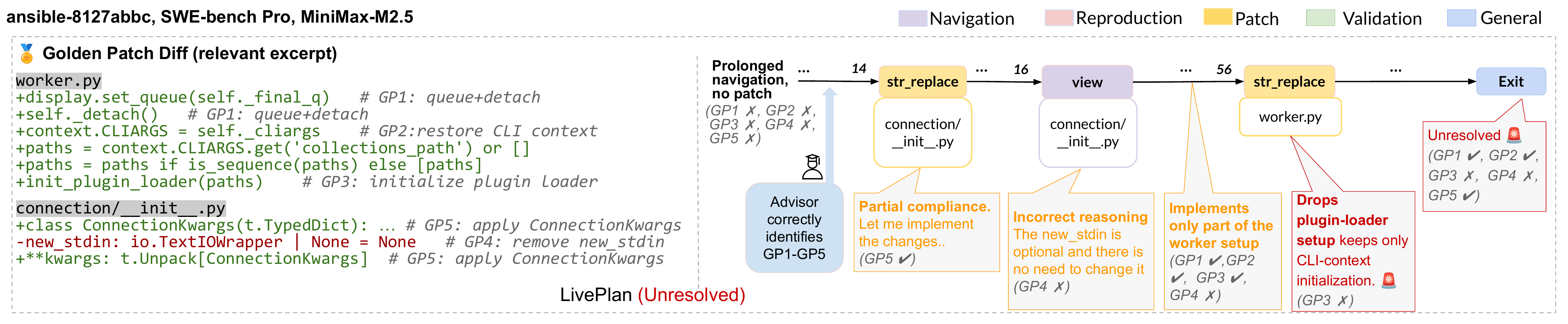}
    \vspace*{-15pt}
    \caption{Case study of a R$\rightarrow$U transition under \approach.}
    \vspace*{-15pt}
    \label{fig:case_study_ru}
\end{figure*}

\approach consistently intervenes before recovery becomes difficult. The first intervention occurs within the first third of the trajectory in all 244 U$\rightarrow$R cases, leaving sufficient opportunity to steer the remaining execution. Figure~\ref{fig:case_study_ur} illustrates such a case (\texttt{\small{ansible-0fd88717}} from \swebp under \Vthree). Both Vanilla and \approach runs initially generate an incomplete patch and then enter validation. As shown in Figure~\ref{fig:case_study_ur}a, the Vanilla run gets stuck in ad-hoc test script creation and execution after several test setup failures, drifting away from the main repair task. 
The trajectory therefore exhibits Prolonged Validation (steps 15--21) and Action Oscillation (self-loop edges at step 18 and 19) before terminating unsuccessfully at step 55.
In contrast, Figure~\ref{fig:case_study_ur}b shows that \approach detects the Prolonged Validation at step 22\footnote{Note that trajectories are not identical and similar behavioral drifts may occur at different steps.} and invokes \advisor, which recommends pausing the ad-hoc tests and instead inspecting and modifying the core logic directly. The executor follows this guidance, edits the correct control flow, performs focused validation, and successfully submits the final patch at step 33. 

\subsubsection{Regressions (R$\rightarrow$U)} 

Regressions across all techniques arise from several sources: \textbf{(1) Misleading or low-quality guidance.} Each method fails differently. \sage may generate an incorrect global plan that redirects the executor toward a non-existent problem (Figure~\ref{fig:sage_example}) \monitoronly is often insufficient for task-specific repair. \periodicadvisor may misdiagnose the trajectory state and generate unhelpful guidance (Figure~\ref{fig:periodic-approach-comparison-example}). \textbf{(2) Execution non-determinism.} The rerun may diverge from Vanilla execution before any intervention, causing failures absent in the original trajectory. \textbf{(3) Brittle Vanilla successes.} Some Vanilla runs succeed by chance despite limited or missing navigation and validation—likely due to memorization/data contamination~\cite{opus_memorization} rather than proper reasoning, making corrective intervention confusing. \textbf{(4) Incomplete compliance.} The executor may fail to faithfully follow or fully implement a correct advice. 
\textbf{(5) Secondary-task distraction.} Advice may intentionally redirect the executor toward secondary issues (e.g., environment configuration), distracting it from the primary repair task.

In nearly all \approach R$\rightarrow$Us, the root cause is not incorrect advice or a failure to intervene on time, but rather the executor's inability to follow the correct guidance. We confirmed the advice's accuracy by referring to the golden patch.
Figure~\ref{fig:case_study_ru} shows such a case (\texttt{\small{ansible-8127abbc}} from \swebp under \minimaxtwo). 
\monitor detects a prolonged navigation without generating a patch and triggers \advisor (step~14). The \advisor correctly identifies all five key repair elements (GP1--GP5), consistent with the gold patch, and directs the executor to begin editing.
The executor follows part of the advice, e.g., modifying GP5 at step~14. Later, it (incorrectly) reasons that the GP3 edit was incorrect, reverting it at step~56, and ultimately submits a partial patch. 
In contrast, the Vanilla run remains in navigation until step~24, but collects all the context to generate a successful patch.  

\noindent \textbf{Guidelines for Future Research.} 
Inference-time intervention should be complemented by intervention-aware post-training. While \approach identifies \emph{when} and \emph{how} to steer execution, successful recovery depends on the executor faithfully carrying out the advice. Future work could train on intervention-conditioned trajectories using process-level rewards that encourage executors to satisfy explicit guidance constraints across following steps, since optimizing only for final task success may encourage brittle, under-validated solutions. Post-training complements this work: \approach determines \emph{when} and \emph{what} advice to give, while the trained executor learns to reliably translate that advice into a correct solution.
Given the execution non-determinism observed here, the evaluation should use repeated paired rollouts to estimate how an intervention changes the probability of resolution, rather than drawing conclusions from a single trajectory. 
\subsection{RQ5: Impact of Non-determinism}
\label{subsec:evaluation-rq5}

\begin{table}[t]
\renewcommand{\arraystretch}{0.9}
\setlength{\aboverulesep}{2pt}
\setlength{\belowrulesep}{2pt}
\caption{Effectiveness in resolving instances of \swebp that deterministically pass or fail in Vanilla runs.}
\label{tab:subset-success-rate}
\setlength{\tabcolsep}{4pt}
\begin{tabular}{cccll}
\toprule
\multicolumn{2}{c}{\textbf{Model}} &
\multicolumn{1}{c}{\multirow[c]{2}{*}{\textbf{\#Instances}}} &
\multicolumn{1}{c}{\multirow[c]{2}{*}{\textbf{Method}}} &
\multicolumn{1}{c}{\multirow[c]{2}{*}{\shortstack{\textbf{Success}\\\textbf{Rate (\%)}}}} \\
\cmidrule(lr){1-2}
\textbf{Advisor} & \textbf{Executor} & & & \\
\midrule
- & \multirow[c]{5}{*}{\shortstack{DeepSeek\\(V3)}} & \multirow[c]{5}{*}{222} & Vanilla & 19.82 \\
GPT-5.2 & & & SAGE & 16.67 {\scriptsize(-3.15)} \\
- & & & \monitoronly & 20.27 {\scriptsize(+0.45)} \\
GPT-5.2 & & & \periodicadvisor & 23.87 {\scriptsize(+4.05)} \\
GPT-5.2 & & & \approach & \textbf{27.03} {\scriptsize(+7.21)} \\
\midrule
- & \multirow[c]{5}{*}{\shortstack{Gemini\\(2.5-flash)}} & \multirow[c]{5}{*}{213} & Vanilla & 8.92 \\
GPT-5.2 & & & SAGE & 11.74 {\scriptsize(+2.82)} \\
- & & & \monitoronly & 12.68 {\scriptsize(+3.76)} \\
GPT-5.2 & & & \periodicadvisor & 18.31 {\scriptsize(+9.39)} \\
GPT-5.2 & & & \approach & \textbf{18.78} {\scriptsize(+9.86)} \\
\midrule
- & \multirow[c]{5}{*}{\shortstack{MiniMax\\(M2.5)}} & \multirow[c]{5}{*}{234} & Vanilla & 52.99 \\
MiniMax-M3 & & & SAGE & 50.00 {\scriptsize(-2.99)} \\
- & & & \monitoronly & 52.99 {\scriptsize(+0.00)} \\
MiniMax-M3 & & & \periodicadvisor & 54.27 {\scriptsize(+1.28)} \\
MiniMax-M3 & & & \approach & \textbf{56.41} {\scriptsize(+3.42)} \\
\bottomrule
\end{tabular}
\end{table}

We re-evaluate all methods on the subset of \swebp with deterministic outcomes across two repeated Vanilla runs.
This reduces the number of instances but retains the majority of instances: 
84\%, 80\%, and 88\% of the study in \S \ref{subsec:evaluation-rq1} for each model.
The distribution of instances with consistent outcomes is similar to that of the complete set ($20.3\%$ easy, $41.0\%$ medium, $36.8\%$ hard, $1.9\%$ very hard): \Vthree ($19.4\%$ easy, $40.1\%$ medium, $38.3\%$ hard, $2.3\%$ very hard); 
\gemini ($21.1\%$ easy, $39.4\%$ medium, $37.6\%$ hard, $1.9\%$ very hard); and \minimaxtwo ($21.8\%$ easy, $40.2\%$ medium, $36.3\%$ hard, $1.7\%$ very hard). 
Because this subset is defined solely by Vanilla's consistency, it cannot selectively favor or penalize any.

As shown in Table~\ref{tab:subset-success-rate}, \approach preserves the method ranking from Table~\ref{tab:success-rate}, achieving the highest success rate across all evaluated models, followed by \periodicadvisor, \monitoronly, and Vanilla. It does so with an advantage comparable in magnitude to that of the full study, indicating that the gain is not an artifact of baseline volatility. We further repeat the \swebv Vanilla runs for \Vthree (limiting this validation to one model to contain experimental cost), where \approach again substantially outperforms Vanilla, improving the success rate from 36.58\% to 46.56\%, a margin consistent with its full-set gain. \textbf{These results confirm that the gains of \approach are robust to execution non-determinism rather than favorable run-to-run variation.}

\section{Related Work}
\label{sec:related-work}

\noindent \textbf{Trajectory analysis and failure diagnosis.} 
Liu et al.~\cite{graphectory} propose process-centric trajectory representations via structured graphs and language abstractions. Chen et al.~\cite{chen2025beyond} and Liu et al.~\cite{liu2025empirical} conduct process-oriented error analysis on GitHub issue-resolution trajectories. TRAIL~\cite{trail} and MAST~\cite{mast} develop failure taxonomies, while FALAT~\cite{rafi2026falat} and AgentRx~\cite{barke2026agentrx} offer step-wise failure localization frameworks. These provide post-hoc analysis but do not intervene during execution; \approach instead monitors execution continuously and intervenes upon detecting drift, correcting behavior before it propagates.

\noindent \textbf{Online steering, planning, and self-correction.} Self-Refine~\cite{madaan_et_al_2023} and LATS~\cite{zhou_et_al_2024} use iterative self-reflection but target stateless reasoning rather than long-horizon agent execution. Planning-based approaches, Plan-and-Act~\cite{erdogan_et_al_2025}, EAGLET~\cite{si_et_al_2025}, and ReCAP~\cite{zhang_et_al_2025}, decouple planning from execution or learn stronger planners. For software engineering agents, SAGE~\cite{sage} regenerates a plan from a completed trajectory and reruns the executor; SWE-PRM~\cite{swe_prm} and Wink~\cite{wink} periodically invoke an LLM evaluator over recent windows; TrajEval~\cite{kim2026trajeval} compares fine-grained trajectories against reference patches. \approach instead avoids periodic LLM judgment and unnecessary interventions while providing targeted next-step advice, achieving the highest resolution rate across all settings, with substantially fewer \advisor interventions.
\section{Threats to the Validity}
\label{sec:threats}

\textbf{External Validity.} 
To ensure the generalizability of the results, we evaluate \approach across three executors and two advisor LLMs on two widely used benchmarks of real-world GitHub issues. We compare \approach with four baselines and ablated versions that resemble related work. To account for execution non-determinism, we further reevaluate \approach on a subset of instances with consistent outcomes across two repeated Vanilla executions. Similar improvements are observed over the Vanilla baseline, suggesting that our approach is robust to run-to-run variation.

\textbf{Internal Validity.}
We investigate the impact of the proposed technique not just on the outcome but also on the process. 

\textbf{Construct Validity.}
Our pipeline is built on peer-reviewed artifacts and validated with well-vetted tools. We analyze the results quantitatively and qualitatively, aligning the two analyses and avoid incorrect metric implementation or measurement.

\section{Concluding Remarks}
\label{sec:conclusion}

We propose \approach for lightweight monitoring and corrective steering of programming agents. \approach relies on two abstract representations of raw trajectories, namely \graph and \lang, to monitor and determine behavioral drifts. It benefits from predefined advice relevant to the drift or custom LLM-generated advice when needed. In the next step, we pursue intervention-aware post-training.

\section{Data Availability}
\label{sec:availability}
The artifacts of this paper are publicly available at ~\cite{website}.

\section*{Acknowledgments}
This work is supported by NSF CCF-2238045 and IBM-Illinois Discovery Accelerator Institute (IIDAI) grants. 

\bibliographystyle{IEEEtranS}  
\bibliography{references}

\begin{thebibliography}{10}
\providecommand{\url}[1]{#1}
\csname url@samestyle\endcsname
\providecommand{\newblock}{\relax}
\providecommand{\bibinfo}[2]{#2}
\providecommand{\BIBentrySTDinterwordspacing}{\spaceskip=0pt\relax}
\providecommand{\BIBentryALTinterwordstretchfactor}{4}
\providecommand{\BIBentryALTinterwordspacing}{\spaceskip=\fontdimen2\font plus
\BIBentryALTinterwordstretchfactor\fontdimen3\font minus \fontdimen4\font\relax}
\providecommand{\BIBforeignlanguage}[2]{{%
\expandafter\ifx\csname l@#1\endcsname\relax
\typeout{** WARNING: IEEEtranS.bst: No hyphenation pattern has been}%
\typeout{** loaded for the language `#1'. Using the pattern for}%
\typeout{** the default language instead.}%
\else
\language=\csname l@#1\endcsname
\fi
#2}}
\providecommand{\BIBdecl}{\relax}
\BIBdecl

\bibitem{opus_memorization}
{Anthropic}, ``Introducing claude opus 4.7,'' \url{https://www.anthropic.com/news/claude-opus-4-7}, 2026.

\bibitem{barke2026agentrx}
S.~Barke, A.~Goyal, A.~Khare, A.~Singh, S.~Nath, and C.~Bansal, ``Agentrx: Diagnosing ai agent failures from execution trajectories,'' \emph{arXiv preprint arXiv:2602.02475}, 2026.

\bibitem{chen2026unlocking}
Y.~Chen, A.~Ahmad, Y.~Zhou, and R.~Jabbarvand, ``Unlocking model potentials through adaptive multi-agent scaffolding for efficient issue resolution,'' \emph{arXiv preprint arXiv:2606.25514}, 2026.

\bibitem{chen2025beyond}
Z.~Chen, W.~Ma, and L.~Jiang, ``Beyond final code: A process-oriented error analysis of software development agents in real-world github scenarios,'' \emph{arXiv preprint arXiv:2503.12374}, 2025.

\bibitem{gemini_2.5_flash}
G.~Comanici, E.~Bieber, M.~Schaekermann, I.~Pasupat, N.~Sachdeva, I.~Dhillon, M.~Blistein, O.~Ram, D.~Zhang, E.~Rosen \emph{et~al.}, ``Gemini 2.5: Pushing the frontier with advanced reasoning, multimodality, long context, and next generation agentic capabilities,'' \emph{arXiv preprint arXiv:2507.06261}, 2025.

\bibitem{deng2025can}
H.~Deng, H.~Zhang, J.~Ou, and C.~Feng, ``Can llm be a good path planner based on prompt engineering? mitigating the hallucination for path planning,'' in \emph{International Conference on Intelligent Computing}.\hskip 1em plus 0.5em minus 0.4em\relax Springer, 2025, pp. 3--15.

\bibitem{swebenchpro}
\BIBentryALTinterwordspacing
X.~Deng, J.~Da, E.~Pan, Y.~Y. He, C.~Ide, K.~Garg, N.~Lauffer, A.~Park, N.~Pasari, C.~Rane, K.~Sampath, M.~Krishnan, S.~Kundurthy, S.~Hendryx, Z.~Wang, C.~B.~C. Zhang, N.~Jacobson, B.~Liu, and B.~Kenstler, ``{SWE-Bench} {Pro}: Can {AI} agents solve long-horizon software engineering tasks?'' Sep. 2025. [Online]. Available: \url{https://arxiv.org/abs/2509.16941}
\BIBentrySTDinterwordspacing

\bibitem{trail}
D.~Deshpande, V.~Gangal, H.~Mehta, J.~Krishnan, A.~Kannappan, and R.~Qian, ``Trail: Trace reasoning and agentic issue localization,'' \emph{arXiv preprint arXiv:2505.08638}, 2025.

\bibitem{erdogan_et_al_2025}
\BIBentryALTinterwordspacing
L.~E. Erdogan, N.~Lee, S.~Kim, S.~Moon, H.~Furuta, G.~Anumanchipalli, K.~Keutzer, and A.~Gholami, ``Plan-and-act: Improving planning of agents for long-horizon tasks,'' in \emph{International Conference on Machine Learning (ICML)}, Jul. 2025. [Online]. Available: \url{https://openreview.net/forum?id=ybA4EcMmUZ}
\BIBentrySTDinterwordspacing

\bibitem{swe_prm}
\BIBentryALTinterwordspacing
S.~Gandhi, J.~Tsay, J.~Ganhotra, K.~Kate, and Y.~Rizk, ``When agents go astray: Course-correcting {SWE} agents with {PRMs},'' in \emph{Workshop on Scaling Environments for Agents (SEA@NeurIPS)}, Dec. 2025. [Online]. Available: \url{https://openreview.net/forum?id=wyrcoDNaGO}
\BIBentrySTDinterwordspacing

\bibitem{sage}
H.~Hayashi, B.~Pang, W.~Zhao, Y.~Liu, A.~Gokul, S.~Bansal, C.~Xiong, S.~Yavuz, and Y.~Zhou, ``Self-abstraction from grounded experience for plan-guided policy refinement,'' \emph{arXiv preprint arXiv:2511.05931}, 2025.

\bibitem{swebench}
\BIBentryALTinterwordspacing
C.~E. Jimenez, J.~Yang, A.~Wettig, S.~Yao, K.~Pei, O.~Press, and K.~R. Narasimhan, ``{SWE}-bench: Can language models resolve real-world github issues?'' in \emph{The Twelfth International Conference on Learning Representations}, 2024. [Online]. Available: \url{https://openreview.net/forum?id=VTF8yNQM66}
\BIBentrySTDinterwordspacing

\bibitem{kim2026trajeval}
M.~Kim, D.~Wang, S.~Cui, F.~Farmahinifarahani, S.~Garg, B.~Ray, T.~Y. Zhuo, R.~Mukherjee, and V.~Kumar, ``Trajeval: Decomposing code agent trajectories for fine-grained diagnosis,'' \emph{arXiv preprint arXiv:2603.24631}, 2026.

\bibitem{lin2025llm}
X.~Lin, Y.~Ning, J.~Zhang, Y.~Dong, Y.~Liu, Y.~Wu, X.~Qi, N.~Sun, Y.~Shang, K.~Wang \emph{et~al.}, ``Llm-based agents suffer from hallucinations: A survey of taxonomy, methods, and directions,'' \emph{arXiv preprint arXiv:2509.18970}, 2025.

\bibitem{deepseek_v3}
A.~Liu, B.~Feng, B.~Xue, B.~Wang, B.~Wu, C.~Lu, C.~Zhao, C.~Deng, C.~Zhang, C.~Ruan \emph{et~al.}, ``Deepseek-v3 technical report,'' \emph{arXiv preprint arXiv:2412.19437}, 2024.

\bibitem{liu2024lost}
\BIBentryALTinterwordspacing
N.~F. Liu, K.~Lin, J.~Hewitt, A.~Paranjape, M.~Bevilacqua, F.~Petroni, and P.~Liang, ``Lost in the middle: How language models use long contexts,'' \emph{Transactions of the Association for Computational Linguistics}, vol.~12, pp. 157--173, 02 2024. [Online]. Available: \url{https://doi.org/10.1162/tacl_a_00638}
\BIBentrySTDinterwordspacing

\bibitem{graphectory}
\BIBentryALTinterwordspacing
S.~Liu, Y.~Chen, R.~Krishna, S.~Sinha, J.~Ganhotra, and R.~Jabbarvand, ``Process-centric analysis of agentic software systems,'' \emph{Proc. ACM Program. Lang.}, vol.~10, no. OOPSLA1, Apr. 2026. [Online]. Available: \url{https://doi.org/10.1145/3798271}
\BIBentrySTDinterwordspacing

\bibitem{plan-compliance}
S.~Liu, S.~Dehghan, J.~Ganhotra, M.~Hirzel, and R.~Jabbarvand, ``Evaluating plan compliance in autonomous programming agents,'' \emph{arXiv preprint arXiv:2604.12147}, 2026.

\bibitem{liu2025empirical}
S.~Liu, F.~Liu, L.~Li, X.~Tan, Y.~Zhu, X.~Lian, and L.~Zhang, ``An empirical study on failures in automated issue solving,'' \emph{arXiv preprint arXiv:2509.13941}, 2025.

\bibitem{liu2026agenthallu}
X.~Liu, X.~Yang, Z.~Li, P.~Li, and R.~He, ``Agenthallu: Benchmarking automated hallucination attribution of llm-based agents,'' \emph{arXiv preprint arXiv:2601.06818}, 2026.

\bibitem{website}
{Liu, Shuyang and Dehghan, Saman and Kim, Jiyoung and Ganhotra, Jatin and Hirzel, Martin and Jabbarvand, Reyhaneh}, ``Artifact repository,'' \url{https://github.com/Intelligent-CAT-Lab/Agent-Planner.git}, 2026.

\bibitem{madaan_et_al_2023}
\BIBentryALTinterwordspacing
A.~Madaan, N.~Tandon, P.~Gupta, S.~Hallinan, L.~Gao, S.~Wiegreffe, U.~Alon, N.~Dziri, S.~Prabhumoye, Y.~Yang, S.~Gupta, B.~P. Majumder, K.~Hermann, S.~Welleck, A.~Yazdanbakhsh, and P.~Clark, ``Self-refine: Iterative refinement with self-feedback,'' in \emph{Conference on Neural Information Processing Systems (NeurIPS)}, Dec. 2023. [Online]. Available: \url{https://proceedings.neurips.cc/paper_files/paper/2023/hash/91edff07232fb1b55a505a9e9f6c0ff3-Abstract-Conference.html}
\BIBentrySTDinterwordspacing

\bibitem{minimax_m2.5}
{MiniMax}, ``Minimax m2.5,'' \url{https://www.minimax.io/models/text}, 2025.

\bibitem{minimax_m3}
------, ``Minimax m3,'' \url{https://www.minimax.io/blog/minimax-m3}, 2026.

\bibitem{wink}
R.~Nanda, C.~Maddila, S.~Jha, E.~M. Khan, M.~Paltenghi, and S.~Chandra, ``Wink: Recovering from misbehaviors in coding agents,'' \emph{arXiv preprint arXiv:2602.17037}, 2026.

\bibitem{gpt_5.2_codex}
{OpenAI}, ``Introducing gpt-5.2-codex,'' \url{https://openai.com/index/introducing-gpt-5-2-codex/}, 2025.

\bibitem{mast}
\BIBentryALTinterwordspacing
M.~Z. Pan, M.~Cemri, L.~A. Agrawal, S.~Yang, B.~Chopra, R.~Tiwari, K.~Keutzer, A.~Parameswaran, K.~Ramchandran, D.~Klein, J.~E. Gonzalez, M.~Zaharia, and I.~Stoica, ``Why do multiagent systems fail?'' in \emph{ICLR 2025 Workshop on Building Trust in Language Models and Applications}, 2025. [Online]. Available: \url{https://openreview.net/forum?id=wM521FqPvI}
\BIBentrySTDinterwordspacing

\bibitem{rafi2026falat}
M.~N. Rafi, M.~Ahasanuzzaman, D.~J. Kim, Z.~Wang, and T.-H. Chen, ``Falat: Tracing failures in llm agent trajectories via dependency-guided search,'' \emph{arXiv preprint arXiv:2606.00765}, 2026.

\bibitem{rao2026autorubric}
D.~Rao and C.~Callison-Burch, ``Autorubric: Unifying rubric-based llm evaluation,'' \emph{arXiv preprint arXiv:2603.00077}, 2026.

\bibitem{sharma2024towards}
M.~Sharma, M.~Tong, T.~Korbak, D.~Duvenaud, A.~Askell, S.~Bowman, E.~Durmus, Z.~Hatfield-Dodds, S.~Johnston, S.~Kravec \emph{et~al.}, ``Towards understanding sycophancy in language models,'' in \emph{International Conference on Learning Representations}, vol. 2024, 2024, pp. 110--144.

\bibitem{shinn2023reflexion}
N.~Shinn, F.~Cassano, A.~Gopinath, K.~Narasimhan, and S.~Yao, ``Reflexion: Language agents with verbal reinforcement learning,'' \emph{Advances in neural information processing systems}, vol.~36, pp. 8634--8652, 2023.

\bibitem{si_et_al_2025}
\BIBentryALTinterwordspacing
S.~Si, H.~Zhao, K.~Luo, G.~Chen, F.~Qi, M.~Zhang, B.~Chang, and M.~Sun, ``A goal without a plan is just a wish: Efficient and effective global planner training for long-horizon agent tasks,'' Oct. 2025. [Online]. Available: \url{https://arxiv.org/abs/2510.05608}
\BIBentrySTDinterwordspacing

\bibitem{swe-agent-config}
{SWE-agent}, ``Swe-agent documentation,'' \url{https://swe-agent.com/latest/usage/batch_mode/}, 2025.

\bibitem{contextrot}
{Teresa Torres}, ``Context rot: Why ai gets worse the longer you chat,'' \url{https://www.producttalk.org/context-rot/}, 2026.

\bibitem{wang2026long}
X.~J. Wang, H.~Bai, Y.~Sun, H.~Wang, S.~Zhang, W.~Hu, M.~Schroder, B.~Mutlu, D.~Song, and R.~D. Nowak, ``The long-horizon task mirage? diagnosing where and why agentic systems break,'' \emph{arXiv preprint arXiv:2604.11978}, 2026.

\bibitem{wilcoxon_test}
\BIBentryALTinterwordspacing
F.~Wilcoxon, ``Individual comparisons by ranking methods,'' \emph{Biometrics Bulletin}, vol.~1, no.~6, pp. 80--83, 1945. [Online]. Available: \url{http://www.jstor.org/stable/3001968}
\BIBentrySTDinterwordspacing

\bibitem{SWE-agent}
J.~Yang, C.~E. Jimenez, A.~Wettig, K.~Lieret, S.~Yao, K.~Narasimhan, and O.~Press, ``Swe-agent: agent-computer interfaces enable automated software engineering,'' in \emph{Proceedings of the 38th International Conference on Neural Information Processing Systems}, ser. NIPS '24.\hskip 1em plus 0.5em minus 0.4em\relax Red Hook, NY, USA: Curran Associates Inc., 2024.

\bibitem{yao2022react}
\BIBentryALTinterwordspacing
S.~Yao, J.~Zhao, D.~Yu, N.~Du, I.~Shafran, K.~R. Narasimhan, and Y.~Cao, ``{ReAct}: Synergizing reasoning and acting in language models,'' in \emph{International Conference on Learning Representations (ICLR)}, May 2023. [Online]. Available: \url{https://openreview.net/forum?id=WE_vluYUL-X}
\BIBentrySTDinterwordspacing

\bibitem{zhang_et_al_2025}
\BIBentryALTinterwordspacing
Z.~Zhang, T.~Chen, W.~Xu, A.~Pentland, and J.~Pei, ``{ReCAP}: Recursive context-aware reasoning and planning for large language model agents,'' in \emph{Conference on Neural Information Processing Systems (NeurIPS)}, Dec. 2025. [Online]. Available: \url{https://openreview.net/forum?id=r2ykUnzuGt}
\BIBentrySTDinterwordspacing

\bibitem{zhou_et_al_2024}
\BIBentryALTinterwordspacing
A.~Zhou, K.~Yan, M.~Shlapentokh-Rothman, H.~Wang, and Y.-X. Wang, ``Language agent tree search unifies reasoning, acting, and planning in language models,'' in \emph{International Conference on Machine Learning (ICML)}, Jul. 2024, pp. 62\,138--62\,160. [Online]. Available: \url{https://proceedings.mlr.press/v235/zhou24r.html}
\BIBentrySTDinterwordspacing

\end{thebibliography}

\end{document}